\documentclass[useAMS,usenatbib]{mnras}
\usepackage{amsmath}
\usepackage{graphicx,txfonts,fleqn,enumerate,color}

\newcommand*  {\diff}     {\mathop{}\!\mathrm{d}}

\renewcommand*{\vec}[1]   {\boldsymbol{#1}}
\newcommand*  {\s}[1]     {\mathsf{#1}}
\newcommand*  {\phm}      {\phantom{-}}
\newcommand*  {\changecolor}{\color{red}}
\newcommand   {\changed}[1]		{{#1}}

\title[Towards time symmetric $N$-body integration]
	  {Towards time symmetric \boldmath $N$-body integration}
\author[Walter Dehnen]
       {Walter Dehnen$^1$\thanks{Email: wd11@le.ac.uk}\\
         Department for Physics \& Astronomy,
         University of Leicester,
         Leicester LE1 7RH}
\date{Accepted .
      Received ;
      }
\pagerange{\pageref{firstpage}--\pageref{lastpage}}
\pubyear{2017}
\begin{document}
\maketitle
\begin{abstract}
	Computational efficiency demands \changed{discretised,} hierarchically organised, \changed{and} individually adaptive time-step sizes (known as the block-step scheme) for the time integration of $N$-body models. However, most existing $N$-body codes adapt individual step sizes in a way that violates time symmetry (and symplecticity), resulting in artificial secular dissipation (and often secular growth of energy errors). Using single-orbit integrations, I investigate various possibilities to reduce or eliminate irreversibility \changed{from} the time stepping scheme. Significant improvements over the standard approach are possible at little extra effort. However, in order to reduce \changed{irreversible step-size changes} to negligible amounts, such as suitable for long-term integrations of planetary systems, more computational effort is needed, while exact time reversibility appears elusive for discretised \changed{individual} step sizes.
\end{abstract}

\label{firstpage}
%%%%%%%%%%%%%%%%%%%%%%%%%%%%%%%%%%%%%%%%%%%%%%%%%%%%%%%%%%%%%%%%%%%%%%%%%%%%%%%%
\section{Introduction}
Astrophysical $N$-body problems typically have a large range of dynamical time scales with factors of $10^{2-4}$ between the shortest and longest orbital times. Consequently, instead of using time steps of (fixed or varying) global size, contemporary $N$-body algorithms advance each particle with time steps of a size which is individually adapted along its trajectory \citep[see also][for a recent review]{DehnenRead2011}. There are two components to such an individual time-stepping method: a \emph{time-step function} and a \emph{time-stepping scheme}. The time-step function (or criterion) returns an appropriate step size given the instantaneous state of a particle's trajectory. The time-stepping scheme, on the other hand, is a method that adapts the individual particle step sizes to follow these time-step criteria as best as possible.

\begin{figure*}
	\hfil
	\includegraphics[width=80mm]{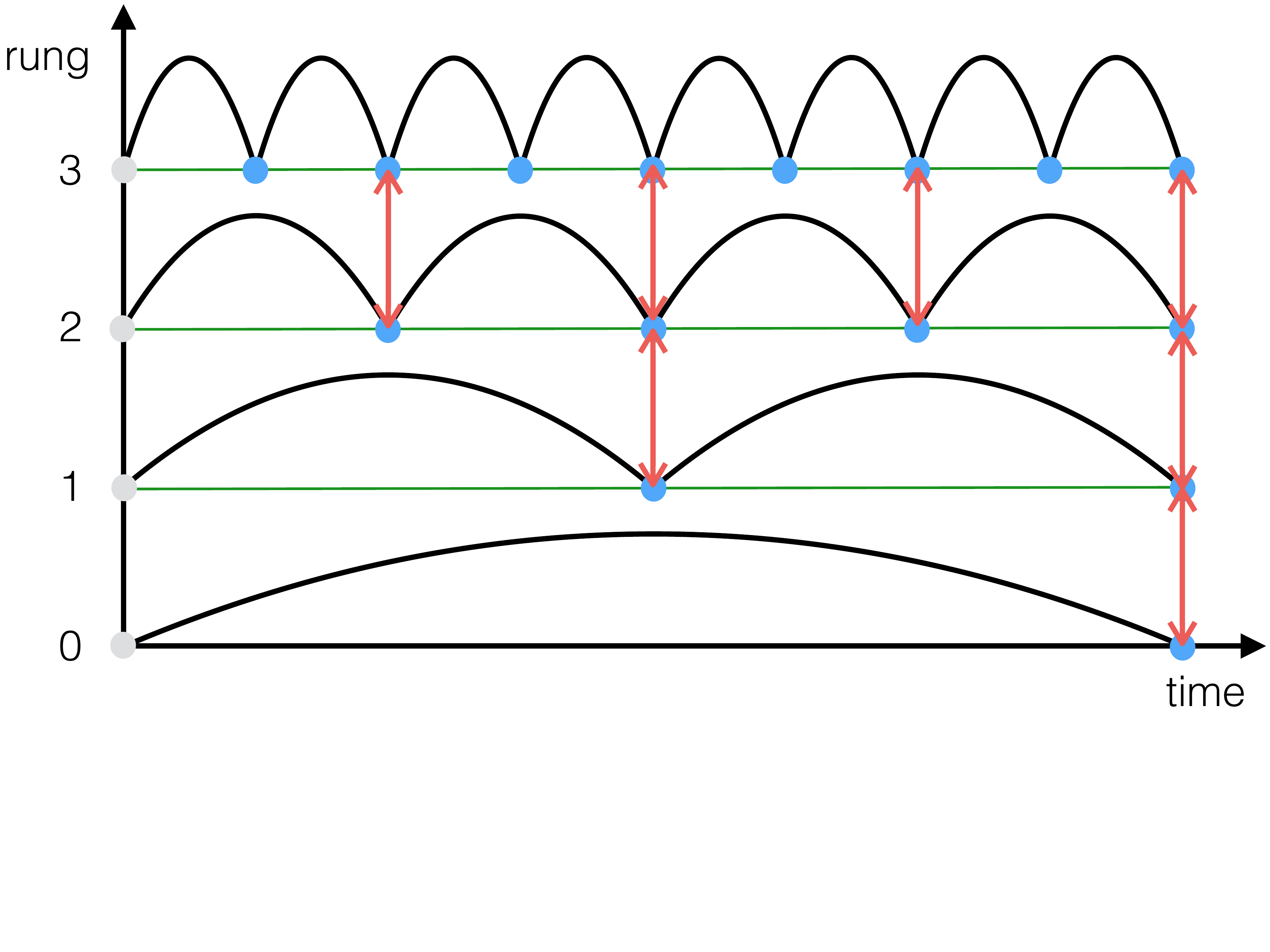}\hspace{10mm}
	\includegraphics[width=80mm]{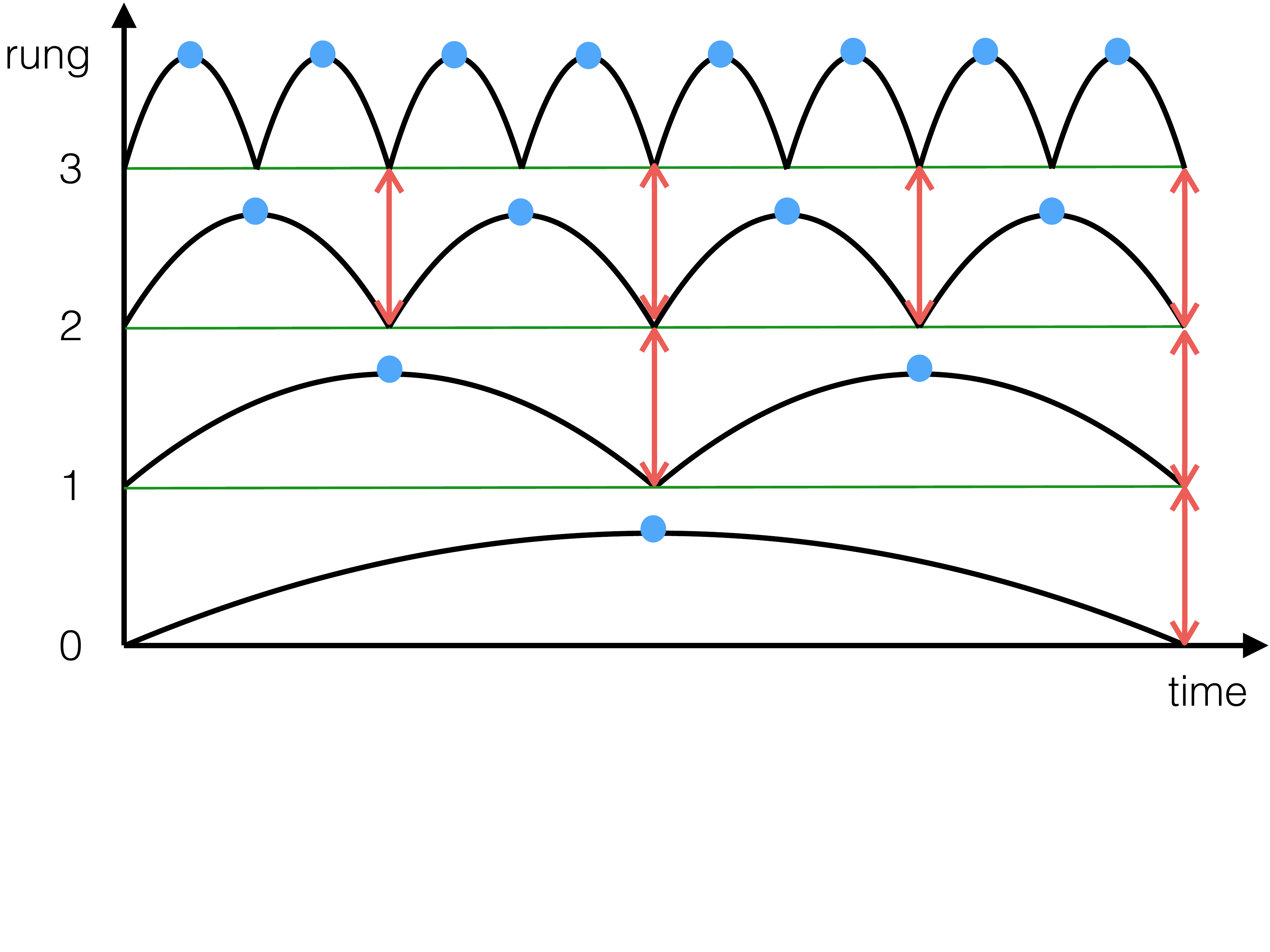}\hfil
	\vspace*{-17mm}
	\caption{\label{fig:block-step}
	Block-step with four rungs for the kick-drift-kick (KDK, left) and the drift-kick-drift (DKD, right) leapfrog. The curves represent individual time steps and the blue dots force computations, when the positions of particles from all rungs need to be predicted. Grey dots indicate forces known from the previous step and red arrows possible changes in rung (step size). The KDK integrator requires half as many predictions as the DKD method and synchronises force calculations not only between particles of the same rung but also between all active rungs.
	}
\end{figure*}

This study is concerned solely with the second ingredient, the time-stepping scheme. There are two important conditions for such a scheme: (1) it should not hinder computational efficiency and (2) it should be time reversible (and/or support symplectic time integration\footnote{A symplectic integrator advances the 
trajectories by a canonical map. An equivalent statement is that the Jacobian
\begin{equation} \label{eq:jacobian}
	\vec{\s{J}} = \frac{\partial\xi(t+h)}{\partial\xi(t)}
\end{equation}
between the initial state $\xi=\{\vec{x},\vec{p}\}$ of the system and that advanced by step size $h$ satisfies $\vec{\s{J}}^T\cdot\vec{\s{\Omega}}\cdot\vec{\s{J}}=\vec{\s{J}}$ with the symplectic matrix
\begin{equation} \label{eq:symplectic}
	\vec{\s{\Omega}} =\begin{pmatrix} \vec{\s{0}}& -\vec{\s{I}} \\ \vec{\s{I}}& \phantom{-}\vec{\s{0}} \end{pmatrix}.
\end{equation}
As a consequence, the geometric structure of phase space, most first integrals, and the Poincar\'{e} invariants are preserved and the energy error tends to be bounded, but see footnote~\ref{note:bad}.}). This latter condition is important to avoid artificial numerical dissipation \citep*{HairerLubichWanner2002}. In order to meet the first condition, all contemporary astrophysical $N$-body methods for large $N$ employ the \emph{block-step} method, where particle time-step sizes are discretised and hierarchically synchronised \citep[][see also Fig.~\ref{fig:block-step} below]{Hayli1967, Sellwood1985, McMillan1986, HernquistKatz1989, Makino1991}. The original motivation for this scheme was the reduction in the number of predictions of particle positions, which are required for the computation of the forces on other particles. Moreover, with modern gravity solvers the \emph{simultaneous} computation of all  gravitational force between $N$ particles requires only $\mathcal{O}(N\ln N)$ \citep[with the tree code, e.g.][]{BarnesHut1986} or (fewer than) $\mathcal{O}(N)$ operations \citep[with the fast multipole method, e.g.][]{Dehnen2000:falcON,Dehnen2002,Dehnen2014}, instead of 
$\mathcal{O}(N^2)$ for a direct force summation, which allows considerable efficiency saving from the synchronisation. It appears that the block-step is the only possibility to achieve these savings and yet allow for individual time-step sizes. It is therefore mandatory to use this method.

%%%%%%%
%%%%%%%
However, (as far as I am aware) none of the contemporary $N$-body codes employing the block-step is time reversible or symplectic, and only little effort has been made towards that goal \citep[with the notable exception of][\changed{see Section~\ref{sec:zonal}}]{Sellwood2014}. One problem is that integrating the mutual forces between any two particles with different step sizes for either particle cannot be reconciled with a canonical map and hence  symplecticity. Note, however, that the block-step method itself does not destroy symplecticity\footnote{\label{note:jump}Any dependence of $h$ on $\xi$ alters the Jacobian~\eqref{eq:jacobian} and destroys symplecticity \citep[e.g.][]{BinneyTremaine2008}. However, when using discrete step sizes (as with the block-step scheme), the function $h(\xi)$ is piece-wise constant with jump discontinuities. Thus, the neighbourhood of almost all trajectories use the same $h$ and the integration remains symplectic (Tremaine, private communication, 2016). The exceptions (trajectories hitting the jumps in $h$) occupy a volume of measure zero in phase space and are irrelevant.}. \citeauthor{FarrBertschinger2007} (\citeyear{FarrBertschinger2007}, see their Fig.~5) demonstrate this with an example where the Poincar\'{e} invariant is conserved to within round-off error using the block-step scheme.

Symplecticity can be restored by integrating the force between any two particles with the same time-step size (either that of the faster or slower of the two particles), resulting in exact conservation of total momentum \citep*{SahaTremaine1994, FarrBertschinger2007, PelupessyJanesPortegiesZwart2012}. However, such schemes are only reversible as long as the time-step adaption process (the application of the jumps discussed in footnote~\ref{note:jump}) is. 

Since the $N$-body dynamics is strictly reversible, irreversibility of the numerical integration method \changed{tends to result} in artificial dissipation. As a consequence, the energy error (for example) is not guaranteed to be bounded, irrespective of whether the integrator is symplectic\footnote{\label{note:bad}Symplecticity by itself does not imply time reversibility. A simple counter example is the second-order accurate leapfrog integrator followed by a positional offset proportional to $h^4$. This is a symplectic, second-order accurate integrator, yet is not reversible and suffers from significant energy drift.} or not. This consideration suggests that reversibility of the integration scheme is more important than symplecticity. Unfortunately, so far no reversible yet efficient block-step-based time-stepping scheme is known. In fact, most practitioners determine the step size by the time-step function evaluated at the start of the time step \citep[e.g.][]{Stadel2001PhDT, Aarseth2003, Springel2005}. This simple forward method violates time symmetry \changed{whenever the step size is adapted} and is well known to destroy long-term stability \citep*[e.g.][]{GladmanDuncanCandy1991, CalvoSanzSerna1993, HairerLubichWanner2002}.
\changed{Thus, the most wanted ingredient for reversible $N$-body integration is a reversible method for adapting the individual particle step sizes given some time-step function.}

The goal of this study is to consider ways to improve this situation. \changed{If $h$ is not discretised, but continuous, the situation is much simpler and various adaptive time stepping methods have been proposed \citep[e.g.][chapter VIII]{HairerLubichWanner2002}. Here, I adapt several of these to discrete $h$ and the block-step and study them in the context of single-orbit integrations. This constitutes a first test that any such method must pass before being considered for the full $N$-body problem.} In particular, I do not toy with \changed{the time-step function or} the underlying time integrator.

The paper is organised as follows. In section~\ref{sec:state} the problem is stated more formally, while sections~\ref{sec:reduce:R:err} and \ref{sec:explicit} present two different approaches of improving reversibility. Section~\ref{sec:diff} presents a third approach, which is more accurate but also more computationally expensive and hence more suitable for long-term integration of planetary systems. Finally, sections~\ref{sec:discuss} and \ref{sec:conclude}discuss the findings and conclude.

%%%%%%%%%%%%%%%%%%%%%%%
%%%%%%%
\begin{figure*}
	\begin{center}
		\includegraphics[width=78mm]{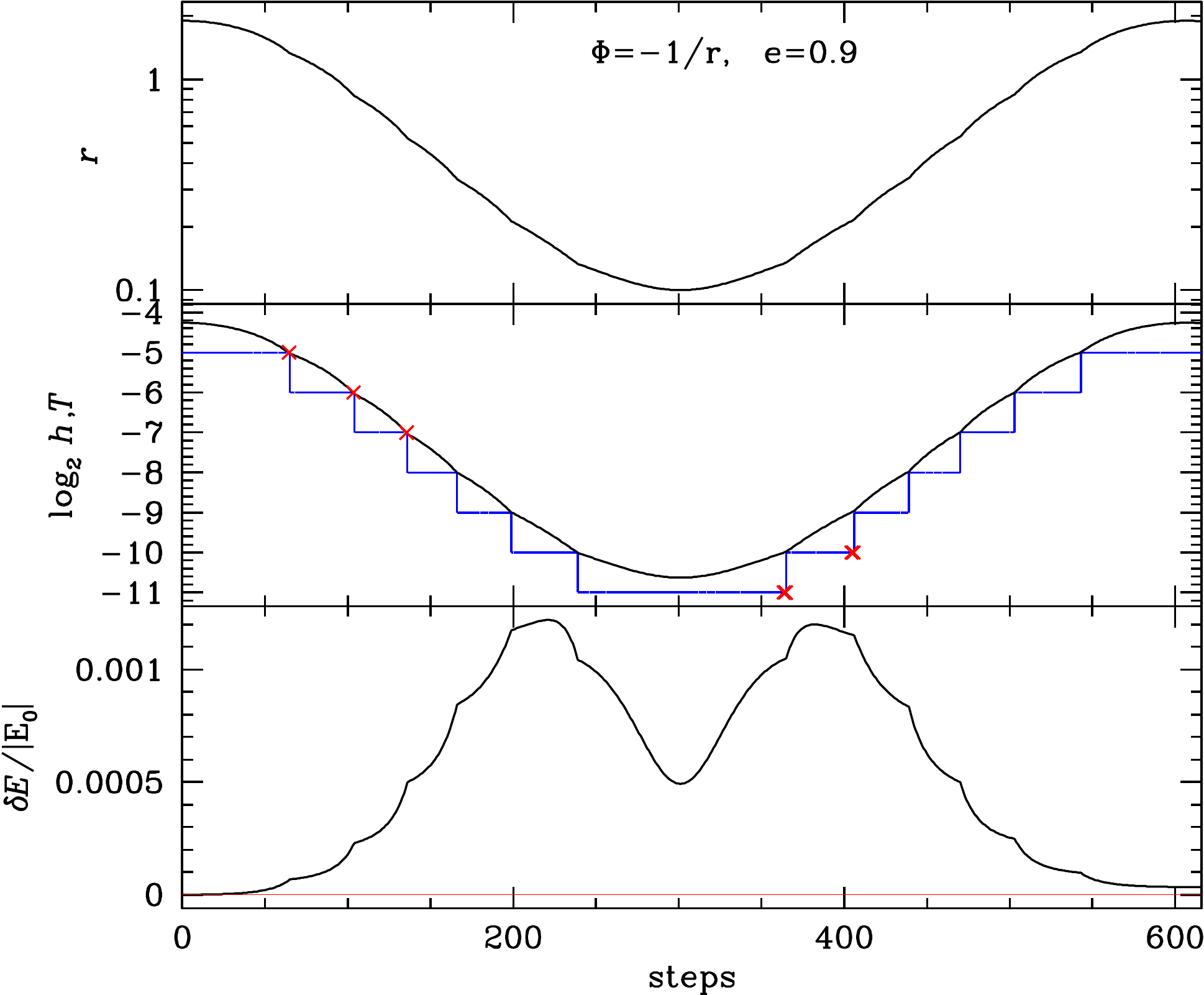}%\\
		\hfil
%		\vspace*{3mm}
		\includegraphics[width=79mm]{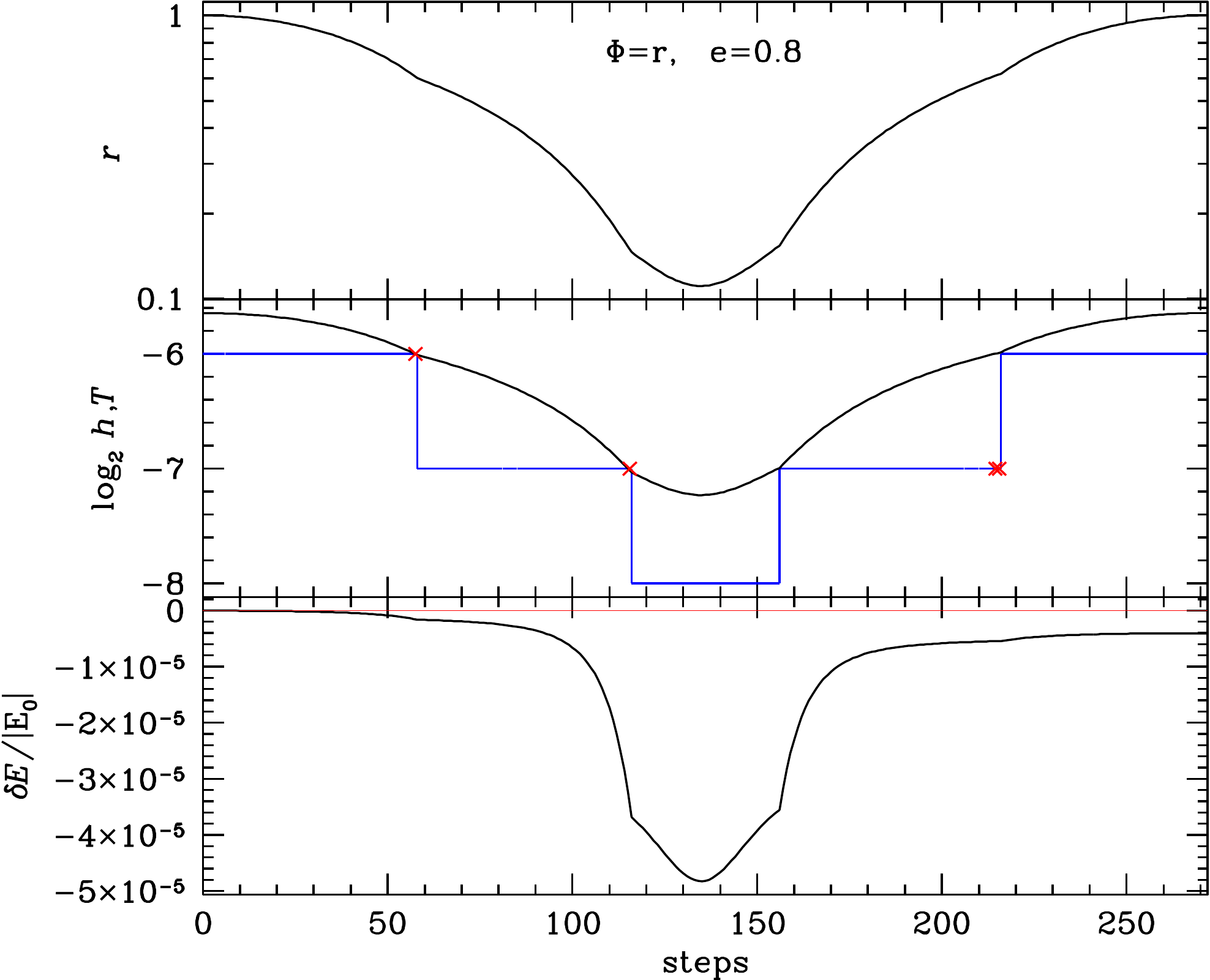}\hspace*{2mm}
	\end{center}
	\caption{\label{fig:fwrd:orbit}
	Radius $r$, step size $h$, stepping function $T$, and relative energy error $\delta E/|E_0|$ as function of time step over one orbital period integrated with the forward scheme~\eqref{eq:fwrd} \changed{using time-step function~\eqref{eq:T:omega} and $\eta=0.02$. The left panels shows an elliptic Kepler orbit with eccentricity $e=0.9$ and the right panels an orbit with $e=0.8$ in a power-law potential, similar to a dark-matter cusp.} Crosses in the middle panels indicate stepping errors of the first two types in equation~\eqref{eq:Rerr} (stepping errors of the other types cannot occur with this scheme)\changed{, which} are responsible for the net energy error. \changed{When integrating the Kepler orbit with the shortest step size used ($h=2^{-11}$), the total number of steps is 10.6 times larger. For the cusp orbit (with $h=2^{-8}$) this ratio is only 2.7.}
	}
\end{figure*}
%%%%%%%
%%%%%%%%%%%%%%%%%%%%%%%%%%%%%%%%%%%%%%%%%%%%%%%%%%%%%%%%%%%%%%%%%%%%%%%%%%%%%%%%
\section{Statement of the problem}
\label{sec:state}
Let $\xi$ denote the state of a particle orbit and $\phi_h$ a \changed{self-adjoint (}time-symmetric\changed{)} integrator used for individual time steps (\changed{i.e.}\ the composition $\phi_{-h}\circ\phi_{h}$ obtains the identity map). Then the state of the integrated orbit after $n$ steps of sizes $h_i$ is
\begin{equation} \label{eq:int:orbit}
	\xi_{n} = \phi_{h_{n-1/2}}\;\xi_{n-1}
		    = \phi_{h_{n-1/2}}\circ\dots\circ\phi_{h_{1/2}}\;\xi_0,
\end{equation}
where $h_{n+1/2}=t_{n+1}-t_n$. Thus, the step sizes naturally have half-integer indices to satisfy time symmetry. Since $\phi_{h}$ is reversible, so is the combined map~\eqref{eq:int:orbit} iff the individual step sizes $h$ are reversibly adapted. With the block-step scheme individual step sizes are discretised as
\begin{equation} \label{eq:h:rung}
	h = 2^{-r} h_{\max}
\end{equation}
with integer \emph{rung} $r\ge0$ and arranged hierarchically, as shown in Fig.~\ref{fig:block-step}. As a consequence of the hierarchical structure, a change to higher rung ($\delta r>0$: shorter step size) is always possible, but a change to lower rung ($\delta r<0$: longer steps size) only if the synchronisation requirement is met (which is every $2^{-\delta r}$ steps). Furthermore, the change in rung is usually limited to $|\delta r|\le1$, corresponding to changing $h$ by a factor two either way. These block-step synchronisation constraints imply that given a desired step size $\tau$, there exist at any time $t_n$ a unique block-step maximum step size
\begin{equation}
	h_{\mathrm{block}}(\tau,t_n) \le \tau
\end{equation}
of the form~\eqref{eq:h:rung}. The dependence of $h_{\mathrm{block}}$ on the actual simulation time $t_n$ originates only from the block-step synchronisation constraint, i.e.\ the fact that doubling the step size (reducing the rung) is not always possible. In view of Noether's theorems, this dependence of the integration method on absolute time suggests that the energy error may never be fully bounded with the block-step method. However, one would expect this time dependence to affect the energy not in a systematic but rather pseudo-random way. In fact, when experimentally dropping the block-step synchronisation constraint in single-orbit integrations, I found if anything a deterioration of the energy conservation. In the notation used in the remainder of this paper, the time-dependence of $h_{\mathrm{block}}$ is suppressed for brevity.

If the integrator $\phi_h$ requires a force at the end (and start) of each time step, then the force computations are synchronised not only between particles with the same rung but between particles of all rungs larger than the smallest active rung at the given time (see also Fig.~\ref{fig:block-step}). This is the main reason why most astrophysical $N$-body methods \changed{for collision-less dynamics} use the kick-drift-kick (rather than the drift-kick-drift) version of the leapfrog integrator
\begin{equation} \label{eq:KDK}
	\phi_h = \phi^{K}_{h/2}\circ\phi^{D}_h\circ\phi^{K}_{h/2}
\end{equation}
with kick and drift operators
\begin{eqnarray}
	\phi^{K}_h\circ\changed{
	\big(t,\,\vec{x},\,\vec{v}\big)} &=&\changed{
	\big(t,\,\vec{x},\,\vec{v}-h\vec{\nabla}\Phi(\vec{x},t)\big)},
%	\begin{pmatrix} t \\ \vec{x} \\ \vec{v} \end{pmatrix} &=&
%	\begin{pmatrix} t \\ \vec{x} \\ \vec{v}-h\vec{\nabla}\Phi(\vec{x},t) \end{pmatrix},
	\\[1ex]
	\phi^{D}_h\circ\changed{
	\big(t,\,\vec{x},\,\vec{v}\big)} &=&\changed{
	\big(t+h,\,\vec{x}+h\vec{v},\,\vec{v}\big)},
%	\begin{pmatrix} t \\ \vec{x} \\ \vec{v} \end{pmatrix} &=&
%	\begin{pmatrix} t +h\\ \vec{x}+h\vec{v} \\ \vec{v} \end{pmatrix},
\end{eqnarray}
where $\vec{x}$ and $\vec{v}$ denote position and velocity\changed{, respectively. All tests in this study are performed using this integrator, but most conclusions regarding the effect of the time-step adaptation schemes are also valid for other integrators, including those of higher order.}

The statement of the problem then is:
\begin{equation}\label{eq:problem}
	\qquad\bullet\quad
	\text{adapt $h$ reversibly and such that $h\approx T(\xi)$},
\end{equation}
where $T(\xi)$ denotes the time-step function. This problem is non-trivial because the states $\xi$, and hence $T(\xi)$, are not synchronised \changed{with the step sizes $h$}, as indicated by the half-integer indices for the \changed{latter}. Thus, in order to solve the problem, some form of synchronisation of relation~\eqref{eq:problem} is required, and different synchronisation attempts result in different stepping schemes.

%%%%%%%%%%%%%%
\subsection{What is wrong with the simple forward method?}
As mentioned in the introduction, the state of the art for adapting individual step sizes is simply
\begin{equation} \label{eq:forward}
	h_{n+1/2} = h_{\mathrm{block}}(T_n)
	\quad\text{with}\quad T_n\equiv T(\xi_n),
\end{equation}
when the step size matches (block-step allowing) \changed{its} optimal \changed{value} at the start of the step. This simply ignores the synchronisation problem in equation~\eqref{eq:problem}. As a consequence, any time dependence of $T$ result in an $\mathrm{O}(h)$ \changed{synchronisation} error \changed{in equation~\eqref{eq:problem}}. The forward method gives the adaptation scheme
\begin{equation} \label{eq:fwrd}
	\delta r
	= \left\{\begin{array}{lllcll}
	-1 \;\;& \text{if} \;\; & h_{n-1/2} & \le & \tfrac{1}{2} T_n & 
	\;\; \text{and block-step allows}, \\[0.5ex]
	\phm 0 & \text{if} & h_{n-1/2} & \le & T_n,
	\\[0.5ex]
	+1 & \multicolumn{4}{l}{\text{otherwise.}}
	\end{array}\right.
\end{equation}
Here, I have limited rung changes to $|\delta r|\le1$, which is common practice and which is assumed in the remainder of this study unless otherwise stated.

\changed{For two orbits integrated with this scheme, Fig.~\ref{fig:fwrd:orbit} plots the time evolution of radius, $h$, $T$, and the energy over one orbital period}. The time-step function \changed{used} is
\begin{equation} \label{eq:T:omega}
	T(\xi) = \eta \sqrt{\frac{r^3}{GM(r)}}
\end{equation}
with $\eta=0.02$ and $M(r)$ the mass enclosed at radius $r$ \changed{(o}f course, such a time-step function is not directly available in $N$-body simulations, but \changed{reasonable} approximations are, e.g. \citealt{ZempEtAl2007}\changed{)}. Evidently, the energy varies considerably over one orbit. Such oscillations of energy are characteristic of symplectic integrators. If using a \changed{(sufficiently small)} fixed step size, the error made on approach to peri-centre is \emph{exactly} undone on the way out again.

%%%%%%%
\begin{figure}
	\hfill\includegraphics[width=84mm]{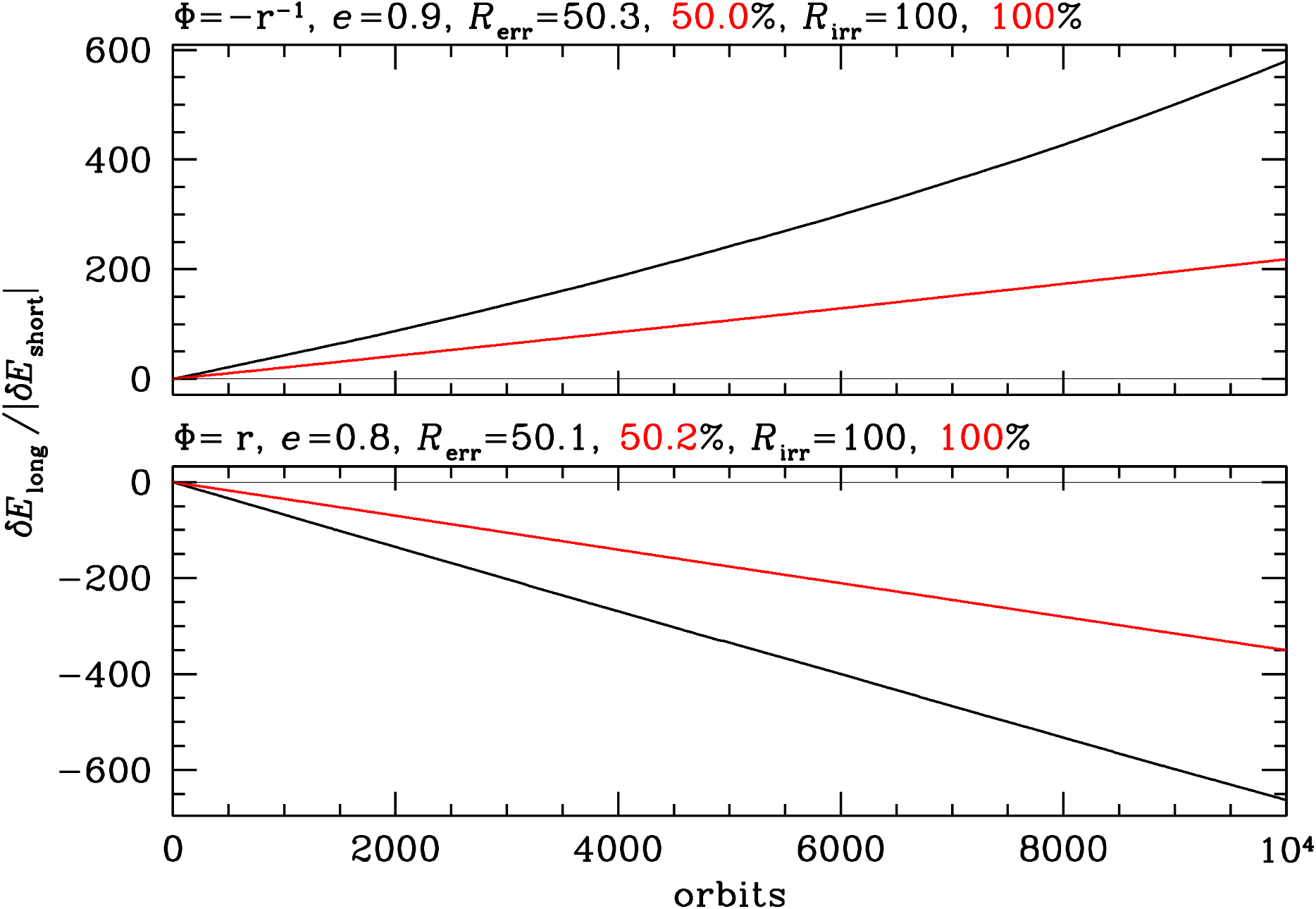}\hfill
	\caption{\label{fig:fwrd:dE}
	\changed{
	Ratio of long- to short-term energy error over $10^4$ periods for the same two orbits as in Fig.~\ref{fig:fwrd:orbit} and $\eta=0.02$ (black) or $\eta=0.01$ (red) integrated with the leapfrog integrator using the state-of-the-art time-step adaptation~\eqref{eq:fwrd}, the `forward scheme'. $\delta E_{\mathrm{long}}$ is the accumulated energy error (measured at apo-centric passages) and $|\delta E_{\mathrm{short}}|$ is the maximum absolute energy error measured over the first period (shown in the bottom panels of Fig.~\ref{fig:fwrd:orbit} for $\eta=0.02$). $R_{\mathrm{err}}$ is the net rate of stepping errors per step-size change, defined in equation~\eqref{eq:Rerr} and $R_{\mathrm{irr}}$ the net rate of irreversible step-size changes, defined in equation~\eqref{eq:Rirr}. With this scheme, every step-size change is irreversible and half of them commit a stepping error.}
	}
\end{figure}
%%%%%%%
With the adaptive method, shorter time steps near peri-centre reduce the energy error there and improve the overall trajectory accuracy\changed{, while avoiding unnecessarily short steps for most of the orbit}. However, the error is not exactly symmetric w.r.t.\ peri-centre. Instead, with the forward adaptation method, the integration accuracy is on average higher \changed{on the outward than on the inward part of the orbit} and a residual error remains. \changed{For
an integrator of order $n$ ($n=2$ for the leapfrog used here), the residual error is proportional to $h^{n+1}\propto\eta^{n+1}$ per irreversible step-size change. As a consequence, the energy error grows over the long term.

This can be seen in Fig.~\ref{fig:fwrd:dE}, which plots  for an integration over $10^4$ periods the ratio of the long-term energy error over the short-term energy error (maximum error over a single orbit as shown in Fig.~\ref{fig:fwrd:orbit}) for $\eta=0.02$ (black) and $0.01$ (red). Evidently, $\delta E_{\mathrm{long}}/|\delta E_{\mathrm{short}}|$ grows roughly linearly at a rate $\propto\eta$. Thus, unless $\eta$ is chosen exceedingly small the long-term error (for these eccentric orbits) is soon dominated by the adverse effects of irreversible step-size changes. This result is independent of the order of the underlying integrator $\phi_h$.}

%%%%%%%%%%%%%%
\subsection{Counting stepping errors}
\label{sec:error}
The departure of the energy error from exactly symmetrical behaviour over each orbit, and hence the net energy error and long-term drift, are \changed{a consequence of each step-size change being irreversible, which results in} deviations of the step size $h$ from the condition~\eqref{eq:problem}. While the forward scheme does not avoid such deviations, one can detect them a posteriori.

There are four types of deviations, depending on the sign of $\dot{T}$ and whether $h$ was too large or too small. One may estimate the ideal step size as $T_{n+1/2}$ if known (i.e.\ for a longer step when in fact two short steps have been taken), otherwise as $\sqrt{T_nT_{n+1}}$. Then a step is too \changed{long} if $h_{n+1/2}>\sqrt{T_nT_{n+1}}$ and \changed{two adjacent steps of the same size $h$ are too short if a larger step was possible and $2h<T_{\mathrm{middle}}$}.

The forward scheme~\eqref{eq:fwrd} only makes two types of errors: too \changed{long} steps at $\dot{T}<0$ and too \changed{short} steps at $\dot{T}>0$. Both tend to result in an energy error of the same sign (which is different for the two orbits considered \changed{in Fig.~\ref{fig:fwrd:orbit}}).

\changed{Errors of} the other types (too \changed{long} steps at $\dot{T}>0$ \changed{or} too \changed{short} steps at $\dot{T}<0$) did not occur, but would have resulted in energy errors of the opposite sign. I define the net rate of stepping errors as difference in the number of errors of different sign divided by the total number of step changes:
\begin{equation}
	\label{eq:R:err}
	\begin{array}{lllll}
	R_{\mathrm{err}} \equiv
	& [ & \text{number of \changed{too long steps} at}  & \text{$\dot{T}<0$} \\
	& + & \text{number of \changed{too short steps} at} & \text{$\dot{T}>0$} \\
	& - & \text{number of \changed{too long steps} at}  & \text{$\dot{T}>0$} \\
	& - & \text{number of \changed{too short steps} at} & \text{$\dot{T}<0$} &
	 \big] / \\
	&   & \multicolumn{2}{l}{\text{total number of step-size changes,}}
	\end{array}
	\label{eq:Rerr}
\end{equation}
\changed{where a pair of adjacent too short steps is only counted once.} \changed{For t}he forward scheme, \changed{the chance for a stepping error is only 50\%, in agreement with the results from the orbit integrations (indicated in Fig.~\ref{fig:fwrd:orbit}).} Of course, $R_{\mathrm{err}}$ is a rough measure, as it only accounts for the typical sign but not the actual size of the errors. In reversed time, $\dot{T}$ changes sign and hence also $R_{\mathrm{err}}$. Therefore, a reversible scheme should obtain $R_{\mathrm{err}}=0$, even though it may not always chose the ideal step size. \changed{Such a scheme should have long-term energy errors comparable to the short-term energy error.}

\changed{
Since the short-term error is $\propto h^n\propto \eta^n$ for an integrator of order $n$, the ratio of long-term errors owed to irreversible step-size changes to the short-term error scales like $\eta R_{\mathrm{err}}$. For the forward method, $R_{\mathrm{err}}\sim0.5$ independent of $\eta$, and therefore $\delta E_{\mathrm{long}}/|\delta E_{\mathrm{short}}|\propto\eta$, in agreement with Fig.~\ref{fig:fwrd:dE}. Of the two orbits presented, the Kepler orbit has three times as many step-size changes than the cusp orbit, but owing to the bimodal structure of the error (bottom left panel of Fig.~\ref{fig:fwrd:orbit}), some of the resulting energy errors cancel, unlike the situation for the cusp orbit. Therefore, the net effect on the ratio $\delta E_{\mathrm{long}}/|\delta E_{\mathrm{short}}|$ happens to be similar for these two orbits.
}

%%%%%%%%%%%%%%%
\subsection{How can the problem be solved?}

%%%%%%%%
\subsubsection{Reducing or eliminating stepping errors}
If one can reduce or even eliminate stepping errors, or at least their net rate such that errors of opposite sign largely cancel, the integration will be near-reversible, even if no attempt is made at constructing an exactly reversible scheme.

One possibility is to extrapolate the time-step function into the future. In section~\ref{sec:xtra}, I consider such a method which has \changed{synchronisation} error $\mathcal{O}(h^3)$, i.e.\ as good as the leapfrog's trajectory error, but still incurs stepping errors.

A more rigorous approach is to attempt to eliminate stepping errors by trying different $h$ and then chosing those satisfying the time-step criteria. Applying this approach to each particle separately ignores the interdependence of particle orbits and time-step functions, but is close enough to reversibility for most practical purposes (see section~\ref{sec:care}). However, this approach requires twice as many force computations as actually used. This overhead can be reduced (but not eliminated) when combined with the extrapolation method, see section~\ref{sec:rjct}.

%%%%%%%%%
\subsubsection{Aiming for reversibility}
\label{sec:prob:reverse}
When instead trying to obtain truly reversible schemes, past and future (step sizes) must be treated symmetrically, which severely restricts our hands in how to use our knowledge about the past. In order to avoid computationally expensive iterations, a reversible method must be explicit, i.e.\ the future step size must be obtained from the current value of the time-step function and the past step size in a reversible way. Since $h$ is discretised, the choice for the future is always between a long step (either equal to or twice the previous step size) and two shorter steps. If the decision for the next step size is based solely on the first of these shorter steps, time symmetry is violated, while the second shorter step is beyond the horizon of prediction (at the moment of the decision). Therefore, \changed{in order to maintain time symmetry,} the decision must be based solely on the merit of the long step\changed{: an explicit scheme must prefer longer steps.} Attempts to obtain such schemes are presented in sections~\ref{sec:explicit} and \ref{sec:diff}.

%%%%%%%%%%%%%%%%%%%%%%%%%%%%%%%%%%%%%%%%%%%%%%%%%%%%%%%%%%%%%%%%%%%%%%%%%%%%%%%%
\section{Extrapolating and iterating}
\label{sec:reduce:R:err}
For the situation \changed{of a continuous} step size $h$, formally exact time symmetry can be obtained by matching \changed{$h$} to some mean of the time-step function $T$ at the start and end of the step:
\footnote{\label{note:irr}
	Matching $h$ instead to the time-step function evaluated in the middle of the step 
	\begin{equation} \label{eq:implicit:middle}
		h_{n+1/2} = T_{n+1/2} \equiv T(\phi_{\frac{1}{2}h_{n+1/2}}\xi_n)
	\end{equation}
	fails to obtain exact time symmetry, because in general
	$\phi_{h/2} \neq \phi_{-h/2}\circ\phi_{h}$, such that $T_{n+1/2}$ obtained in the forward and backward directions differ.}
\begin{equation} \label{eq:implicit:mean}
	h_{n+1/2}=\mu\big(T_n,T_{n+1}\big)
	\quad\text{with}\quad
	T_{n+1} \equiv T(\phi_{h_{n+1/2}}\circ\xi_n).
\end{equation}
Here, $\mu(x,y)$ denotes a general mean, satisfying $\mu(x,y)=\mu(y,x)$ and $\min\{x,y\}\le\mu(x,y)\le\max\{x,y\}$, for example the arithmetic mean
\begin{equation} \label{eq:implicit:ave}
	h_{n+1/2}=\tfrac{1}{2}\big(T_n+T_{n+1}\big)
\end{equation}
\citep*{HutMakinoMcMillan1995}, the geometric mean
\begin{equation} \label{eq:implicit:geo}
	h_{n+1/2}=\sqrt{T_nT_{n+1}},
\end{equation}
or the harmonic mean. Equation~\eqref{eq:implicit:mean} is a non-trivial implicit relation for $h_{n+1/2}$, which requires an iterative approach for an exact solution.

\changed{For $h$ discretised with} the block-step, equation~\eqref{eq:implicit:mean} is naturally replaced by 
\begin{equation} \label{eq:implicit:mean:block}
	h_{n+1/2}=h_{\mathrm{block}}\Big(\mu\big(T_n,T_{n+1}\big)\Big),
\end{equation}
when the step size can only take a few discrete values, such that a finite number of iterations suffices for full convergence. However, owing to the interdependence of the particle trajectories, an exact solution requires iterating not just over the trajectory of each particle individually, but over the $N$-body trajectory of all particles combined. While this appears to give a reversible scheme \citep{MakinoEtAl2006}, it requires an enormous computational effort both in time and memory, and is completely unpractical.

%%%%%%%
\begin{figure}
	\hfill\includegraphics[width=84mm]{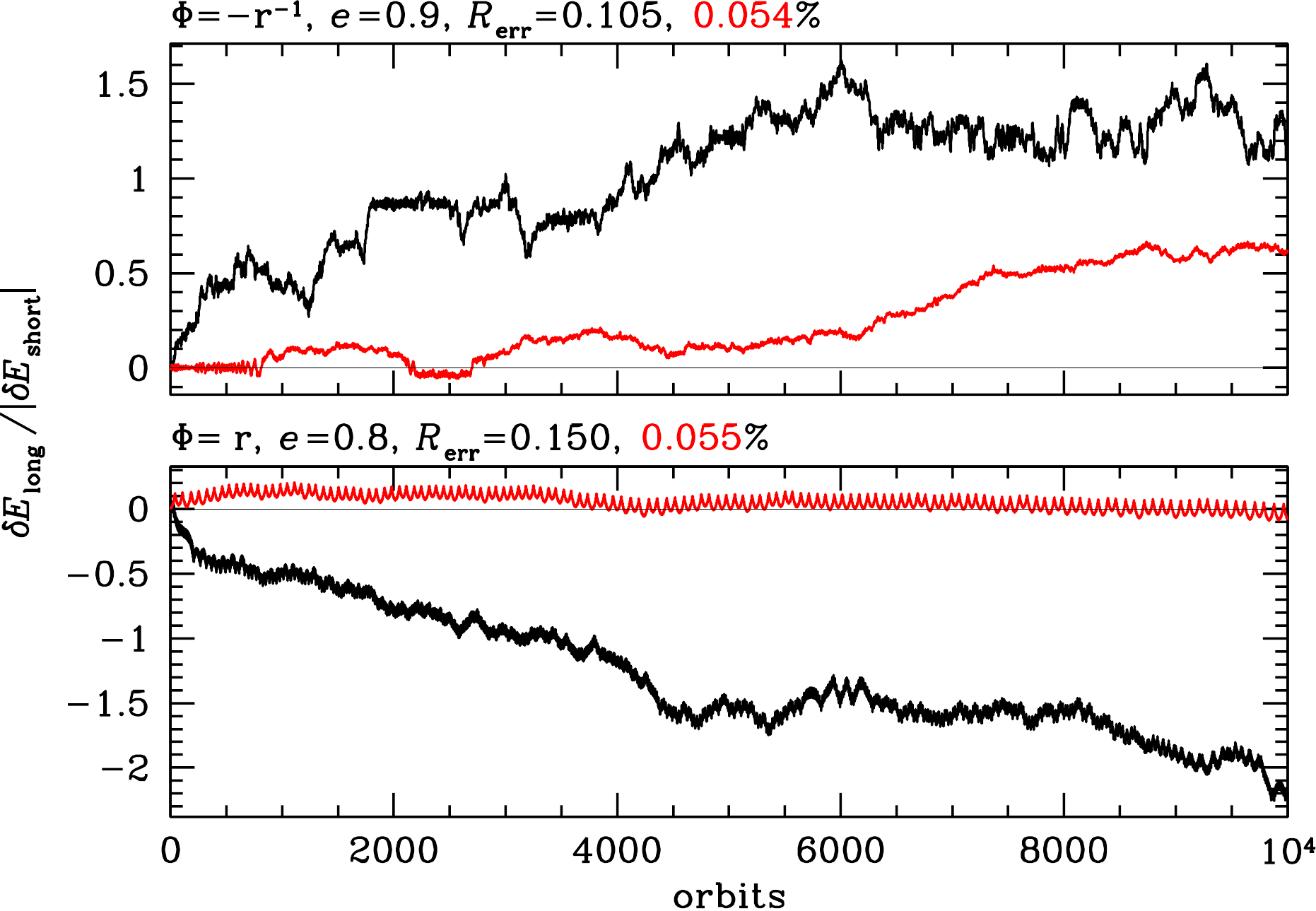}\hfill
	\caption{\label{fig:xtra:dE}
	As Fig.~\ref{fig:fwrd:dE} but for the scheme~\eqref{eq:xtra} of Section~\ref{sec:xtra}. \changed{Note that orbital fluctuations of the energy error (which have amplitude of 1 in this plot) are suppressed by measuring it only at apo-centre, i.e.\ at the same phase for each orbit.}
	}
\end{figure}
%%%%%%%
%%%%%%%%%%%%%%%%%%
\subsection{Extrapolating the time-step function}
\label{sec:xtra}
One possibility to avoid such iterations is to \emph{estimate} $T_{n+1}$ and obtain an approximately reversible scheme. The lowest-order approximation is simply $T_{n+1}\approx T_n$, when equation~\eqref{eq:implicit:mean} results in the forward scheme~\eqref{eq:forward}. At the next order, one can use the previous value of $T$ and estimate
\begin{equation} \label{eq:T:linear}
	\sqrt{T_{n}T_{n+1}} \approx 
	T_n \left(\frac{T_n}{T_{n-1}}\right)^{\frac{\phantom{2}\,h_{n+1/2}}{2\,h_{n-1/2}}},
\end{equation}
which amounts to linear extrapolation of $\ln(T)$ in time. When inserted into equation~\eqref{eq:implicit:geo} this still gives an implicit relation for $h_{n+1/2}$, but with a trivial dependence. For the block-step with $|\delta r|\le1$, $h_{n+1/2}$ can take only three allowed values and one can easily solve this relation to obtain the scheme
\begin{equation} \label{eq:xtra}
	\delta r = \left\{
	\begin{array}{llcl}
		-1 \; & \text{if}\;\; h_{n-1/2} T_{n-1} &\le& \tfrac{1}{2}T_n^2
		\;\; \text{and block-step allows}, \\[0.5ex]
		\phm0 & \text{if}\;\; h_{n-1/2}^2 T_{n-1} &\le& T_n^3,
		\\[0.5ex]
		+1 & \text{otherwise.}
	\end{array}\right.
\end{equation}
Strictly, these conditions are not unique if $h_{n-1/2}\ge2T_n$. While this should never occur for appropriate time-step functions, it can be easily resolved by testing for longer steps first, say.

Equation~\eqref{eq:T:linear} has an error $\propto h^2 (\diff^2\ln(T)/\diff t^2)$, but since even-order time derivatives are time symmetric, the \changed{synchronisation} error of scheme~\eqref{eq:xtra} is $\mathcal{O}(h^3)$ \changed{as opposed to $\mathcal{O}(h)$ for the forward scheme}.

Fig.~\ref{fig:xtra:dE} shows the \changed{ratio of long- to short-term} energy error over $10^4$ periods with this scheme for exactly the same two orbits already considered above with the naive forward scheme. The net rate \changed{$R_{\mathrm{err}}$} of stepping errors \changed{is much} smaller than for the forward scheme~\eqref{eq:fwrd}, representing a significant improvement.
\changed{As a result}, the \changed{long-term and orbital} energy variations are comparable \changed{in stark contrast to the situation for the forward scheme} (see Fig.~\ref{fig:fwrd:orbit}). \changed{Another significant improvement over the forward method is that $R_{\mathrm{err}}$ decreases with $\eta$. As a consequence, the ratio of irreversibility-induced long-term energy errors to the orbital energy error decreases faster than $\propto\eta$.

In this context, it is worth noting that not all of the measured long-term energy error is due to irreversible step-size changes (it also occurs for integrations without any violation of time symmetry, such as for the Kepler orbit integrated with $\eta=0.01$ in Fig.~\ref{fig:intg:dE} below). As already discussed in section~\ref{sec:state}, Noether's theorems and the dependence of $h_{\mathrm{block}}(\tau)$ on absolute time suggest such pseudo-random behaviour of the long-term energy error. Alternatively, one may interpret this as a random walk caused by an incommensurability between period and time stepping (the cusp orbit exhibits some quasi-periodic behaviour of the error, clearly visible at $\eta=0.01$ in Figs.~\ref{fig:xtra:dE}-\ref{fig:rjct:dE}, \ref{fig:symm:dE:mid} \& \ref{fig:intg:dE}, indicating a resonance between period and time stepping).
}

%%%%%%%%%%%%%%%%%%
\subsection{A try-and-reject scheme}
\label{sec:care}

The extrapolation of the time-step function is only advisable if it is well-behaved, i.e.\ does not change much on time scales of itself, implying $|\dot{T}|\ll1$. This is not necessarily the case in $N$-body simulations, where fluctuations may occur due to close and insufficiently softened encounters as well as force approximation errors.

Therefore, I now consider an alternative approach to deal with the implicit relations. If one ignores the violation of time symmetry resulting from the mutual dependence of the particle trajectories, one can consider each particle trajectory separately. Since there are only few discrete values allowed for $h_{n+1/2}$, one can, instead of iterating, simply try them one after the other, starting with the shortest allowed step.

To this end, the position at $t_n+\tfrac{1}{2}h_{n-1/2}$, corresponding to halving the step size, is predicted first and the time-step function $T$ at that point computed. If $h_{n-1/2}>T$ the step size $h_{n+1/2}=h_{n-1/2}$ is too long (in the sense of equation~\ref{eq:implicit:middle}) and instead the shorter step must be used, the first part of which has already be done by predicting the position. Otherwise, the step is continued to size $h_{n+1/2}=h_{n-1/2}$ at the end of which the procedure is repeated (subject to block-step synchronisation constraints).

In order to continue the step, the kick-drift-kick leapfrog can be re-phrased as a predictor-corrector scheme. The predictor simply predicts position and velocity into the future assuming a constant acceleration $\vec{a}=\vec{a}_{\mathrm{beg}}$ equal to that at the start  of the step. The prediction operation is associative and second-order accurate in position. A full time step consists of any number of predictions followed by a force computation (obtaining $\vec{a}_{\mathrm{end}}$) and the correction step
\begin{equation} \label{eq:correct}
	\vec{v} \leftarrow\vec{v} + \tfrac{1}{2} h\;
			(\vec{a}_{\mathrm{end}}-\vec{a}_{\mathrm{beg}}).
\end{equation}
In practice, this scheme requires storage for $\vec{a}_{\mathrm{beg}}$ and $h$, the accumulated length of the step so far (which is incremented by the predictor). The scheme for a full time step for one particle is then as follows. 
\begin{itemize}
	\item[(i)] Increment the rung ($r\leftarrow r+1$) to try a shorter step.	\item[(ii)] \label{scheme:predict}
		Predict $\xi$ until the rung $r$ is block-step synchronised.
	\item[(iii)] Calculate $T$ and the force (obtaining $\vec{a}_{\mathrm{end}}$)\footnote{Strictly speaking, $\vec{a}_{\mathrm{end}}$ is only required in (v), but since $T$ typically depends on the gravitational potential and/or acceleration, the efficiency saving by evaluating $\vec{a}_{\mathrm{end}}$ only when needed is not available.}.
	\item[(iv)] If $2h\le T$ and if rung $r-1$ is block-step synchronised at time $t+h$, reject the step (discarding the force computation just made), decrement the rung ($r\leftarrow r-1$), and recurse with point (ii).
	\item[(v)] Otherwise, finish the step by applying the correction~\eqref{eq:correct}, then reset $\vec{a}_{\mathrm{beg}}\leftarrow\vec{a}_{\mathrm{end}}$ and $h\leftarrow0$. 
\end{itemize}

%%%%%%%
\begin{figure}
	\hfill\includegraphics[width=84mm]{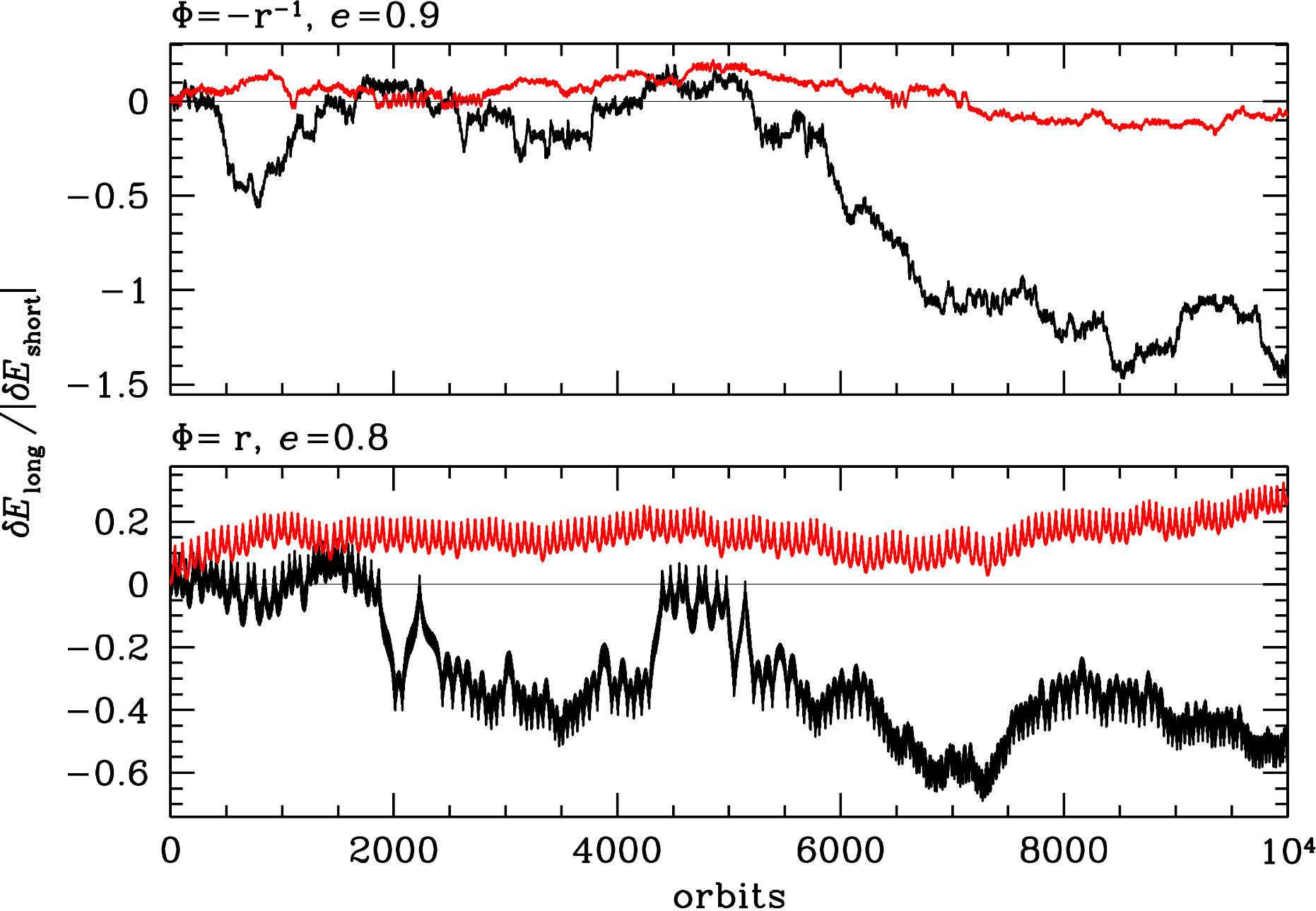}\hfill
	\caption{\label{fig:care:dE}
	As Figs.~\ref{fig:fwrd:dE} \changed{and~\ref{fig:xtra:dE}} but for the scheme of section~\ref{sec:care}. This scheme avoids stepping errors (such that $R_{\mathrm{err}}=0$) at the price of doubling the computational costs.
	}
\end{figure}
%%%%%%%
This scheme is similar in spirit to that proposed by \cite{QuinnEtAl1997}, with the main difference that their scheme tests for longer steps first, while I test for shorter steps first. I should also note that this scheme violates time symmetry not only because of the mutual dependence of particle step sizes, but even when applied to a single orbit in a fixed potential. This is because the predictor is only time symmetric between the start and end of a step, but not with respect to intermediate times. The position predicted for the middle of a step (at unchanged step size) differs between forward and backward steps by $\tfrac{1}{8}h^2(\vec{a}_{\mathrm{end}}-\vec{a}_{\mathrm{beg}})$, see also footnote~\ref{note:irr}. Consequently, the time-step function $T$ evaluated at these positions  in (iii) differs too and with it possibly the decision about the step size in (iv).

However, these violations of time symmetry are not obviously systematic and do not introduce a coherent arrow of time. Hence, no \changed{significant} systematic energy drift is to be expected. \changed{Fig.~\ref{fig:care:dE} shows that for the same two orbits used before the ratio of long- to short-term energy error is slightly smaller than for the extrapolation scheme form the previous sub-section.}

%%%%%%%%%%%%%%%%%%%%%%%%
\subsection{Combining extrapolation with try-and-reject}
\label{sec:rjct}
%%%%%%%
\begin{figure}
	\hfill\includegraphics[width=84mm]{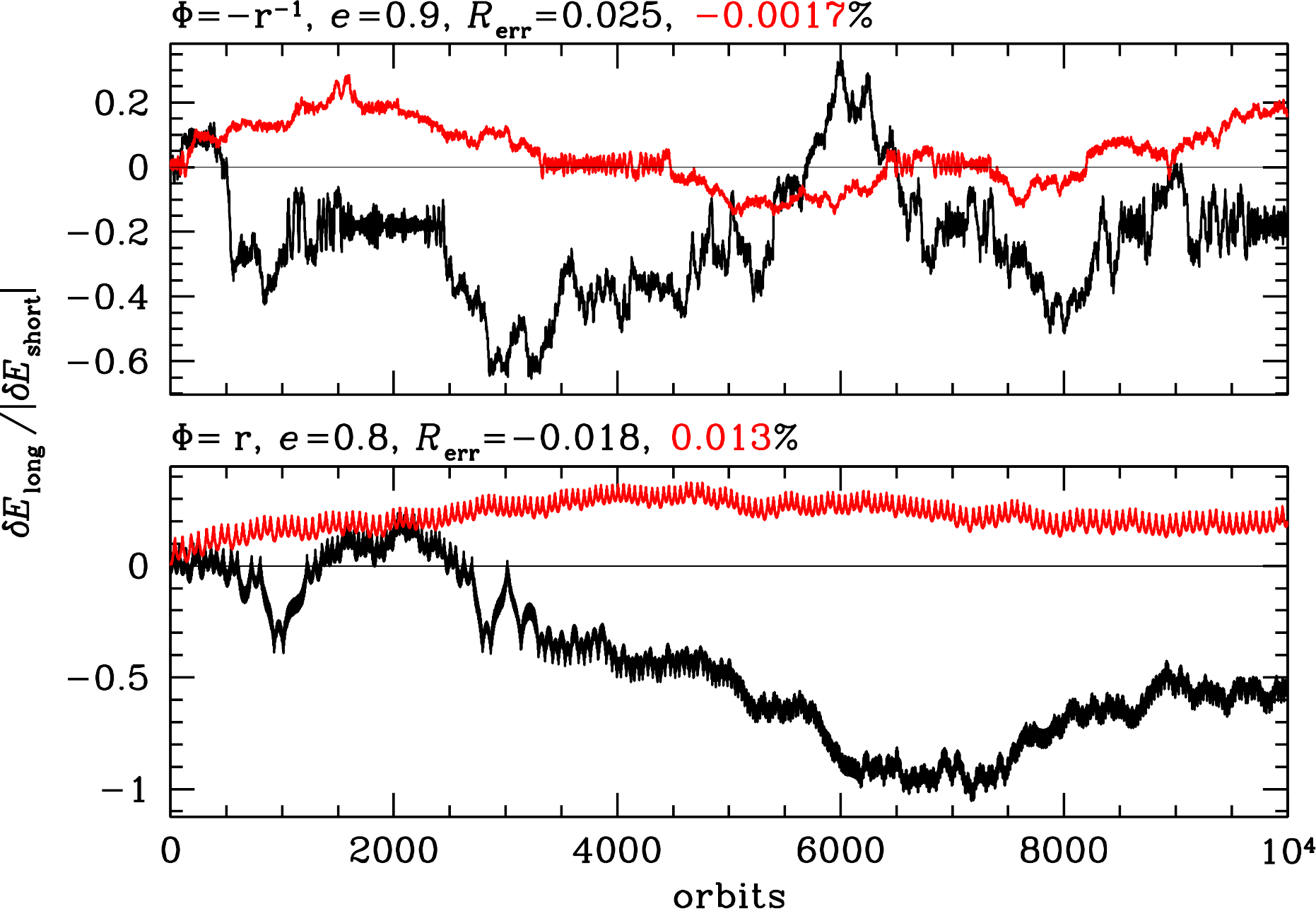}\hfill
	\caption{\label{fig:rjct:dE}
	As Fig.~\ref{fig:fwrd:dE} but for the scheme~\eqref{eq:rjct} using $\lambda=0.8$.
	}
\end{figure}
%%%%%%%
The \changed{try-and-reject} scheme is rather wasteful, as it requires (on average) twice as many force computations as it actually uses for the orbit integration. However, by combining extrapolation of the time-step function with the rejection idea, one \changed{may} avoid to always try for a shorter step. One can parameterise the ignorance about the precise future values for $T$ by a parameter $\lambda\le1$ and assume that
\begin{equation} \label{eq:T:extra:0:lambda}
	T_{n+1/2} \ge T_n\,\lambda^{\frac{\phantom{2}\,h_{n+1/2}}{2\,h_{n-1/2}}},
\end{equation}
which corresponds to the assumption that $\dot{T}\gtrsim\ln(\lambda)$. With this  
supposition, one can replace step (i) in the algorithm given in the previous sub-section by $r \leftarrow r + \delta r$ with
\begin{equation} \label{eq:rjct}
	\delta r = \left\{
	\begin{array}{ll}
		-1 \; & \text{if}\;\; h_{n-1/2} \le \tfrac{1}{2}T_{\!n}\,\lambda
		\;\; \text{and block-step allows}, \\[0.5ex]
		\phantom{-}0  & \text{if}\;\; h_{n-1/2}^2 \le T_n^2\lambda,
		\\[0.5ex]
		\phantom{-}1 & \text{otherwise.}
	\end{array}\right.
\end{equation}
For $\lambda=0.8$, Fig.~\ref{fig:rjct:dE} shows the evolution of the \changed{ratio of long- to short-term} energy error for the same orbits as considered before. \changed{The long-term error is similar to that for} the try-and-reject scheme before (Fig.~\ref{fig:care:dE}). For these orbits, the ratio of force calculations to time steps is down from 2 for the pure try-and-reject scheme to 1.19 and 1.3\changed{6} for Kepler and cusp orbit, respectively.

%%%%%%%%%%%%%%%%%%%%%%%%%%%%%%%%%%%%%%%%%%%%%%%%%%%%%%%%%%%%%%%%%%%%%%%%%%%%%%%%
\section{Attempting explicit reversibility}
\label{sec:explicit}
The schemes discussed in the previous section attempt to obtain $T_{n+1/2}$, i.e.\ synchronise the right-hand side of equation~\eqref{eq:problem}. Instead, the schemes \changed{considered in this section} attempt to synchronise the left-hand side of equation~\eqref{eq:problem}. For continuous step sizes (unconstrained by the blockstep scheme), this can be accomplished by swapping the roles of $h$ and $T$ in the implicit relation~\eqref{eq:implicit:mean}, yielding \citep*{HuangLeimkuhler1997, HolderLeimkuhlerReich2001, LeimkuhlerReich2005}
\begin{equation} \label{eq:explicit:mu}
	\mu(h_{n-1/2},h_{n+1/2}) = T_n,
\end{equation}
which is an explicit relation for $h_{n+1/2}$, i.e.\ requires no iterations for an exact solution. Since in reversed time $h_{n-1/2}$ and $h_{n+1/2}$ are swapped while $T_n$ remains unaltered, this scheme is \changed{reversible}. \citeauthor{HolderLeimkuhlerReich2001} proposed the harmonic mean, $2/\mu=1/x+1/y$. I will use the geometric mean $\mu^2=xy$, which of the one-parameter family of averages
\begin{equation} \label{eq:mu:alpha}
	\begin{array}{rcll}
		\mu^\alpha &=& \tfrac{1}{2}[x^\alpha+y^\alpha]
		\qquad & \text{for $\alpha\neq0$},		\\[1.2ex]
		\ln\mu &=& \tfrac{1}{2}[\ln x+\ln y]
		\qquad & \text{for $\alpha=0$}
	\end{array}
\end{equation}
is the only one that always obtains a unique positive $h_{n+1/2}$ for any given $T_n,\;h_{n-1/2}>0$. In the remainder of this study, I stick to the geometric mean, but I experimented with various $\alpha$ without finding $\alpha=0$ inferior.

The resulting integration schemes \changed{for continuous $h$} give \changed{long-term} energy errors consistent with zero for the two test orbits used before (not shown). It seems expedient, therefore, to try to transfer this idea to the block-step scheme.

For the further quantification of (the lack of) reversibility, I introduce the net rate of irreversible step-size changes in close analogy to the net rate of stepping errors:
\begin{equation}
	\label{eq:R:irr}
	\begin{array}{llllll}
	R_{\mathrm{irr}} \equiv
	& [ & \text{number of steps with} & h>h_{\mathrm{back}}
		& \text{at $\dot{T}<0$} \\
	& + & \text{number of steps with} & h<h_{\mathrm{back}}
		& \text{at $\dot{T}>0$} \\
	& - & \text{number of steps with} & h>h_{\mathrm{back}}
		& \text{at $\dot{T}>0$} \\
	& - & \text{number of steps with} & h<h_{\mathrm{back}}
		&\text{at $\dot{T}<0$} & \big] / \\
	&   & \multicolumn{4}{l}{\text{number of step-size changes.}}
	\end{array}
	\label{eq:Rirr}
\end{equation}
Here, $h_{\mathrm{back}}$ is the step size which the same scheme would have taken when applied after the step of size $h$ in the backward direction. Like stepping errors, irreversibilities come in four flavours, depending on the sign of $\dot{T}$ and the relation between $h$ and $h_{\mathrm{back}}$. Note that $R_{\mathrm{irr}}$ is an exact measure, unlike $R_{\mathrm{err}}$ of equation~\eqref{eq:R:err}, which involved an estimation for the correct step size.

%%%%%%%
\begin{figure}
	\hfill\includegraphics[width=84mm]{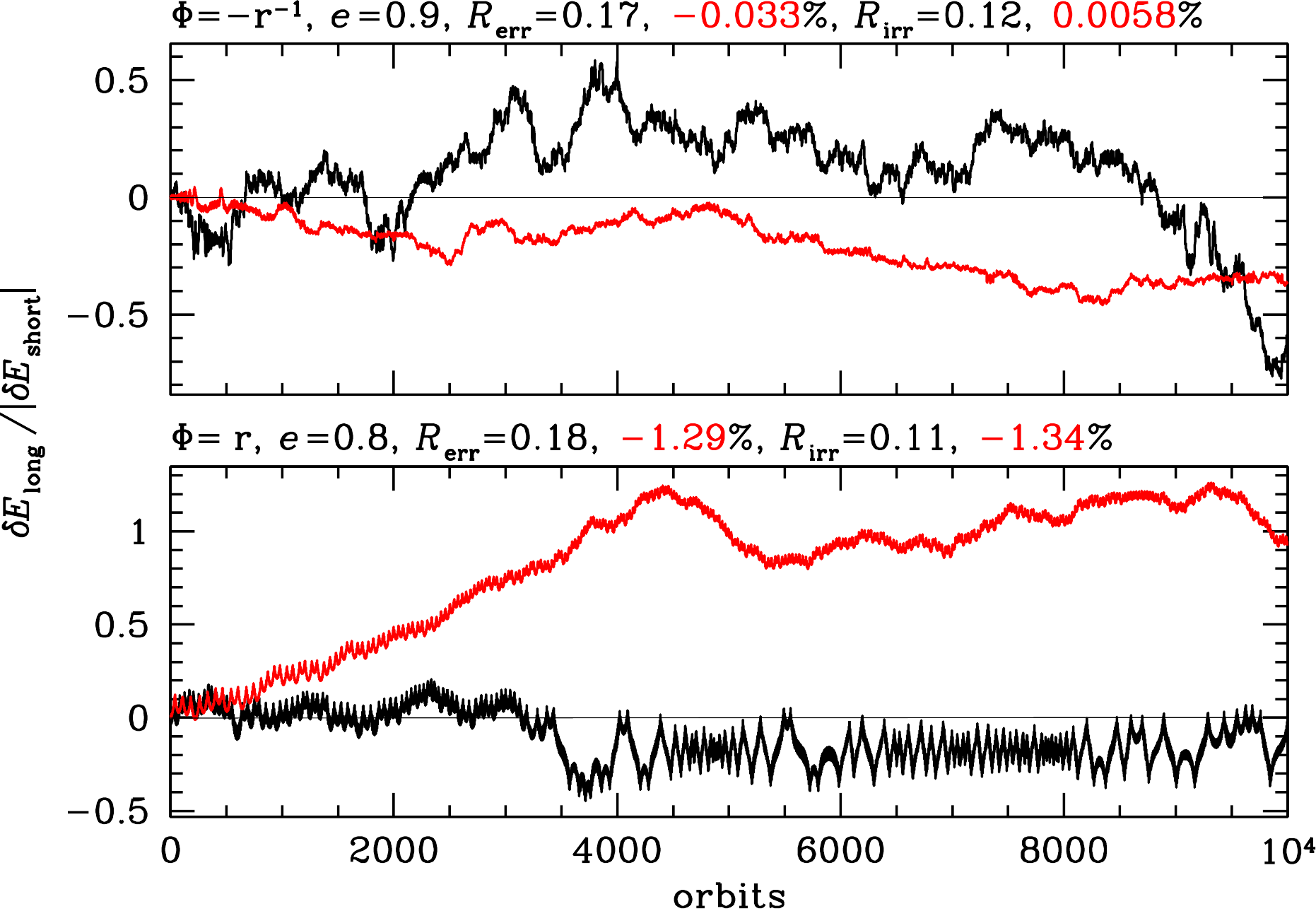}\hfill
	\caption{\label{fig:symm:dE:apo}
	As Fig.~\ref{fig:fwrd:dE} but for the scheme~\eqref{eq:symm} with the orbits started at apo-centre and with initial $\tau_{1/2}=T(0)$.
	}
\end{figure}
%%%%%%%
%%%%%%%
\begin{figure}
	\hfill\includegraphics[width=84mm]{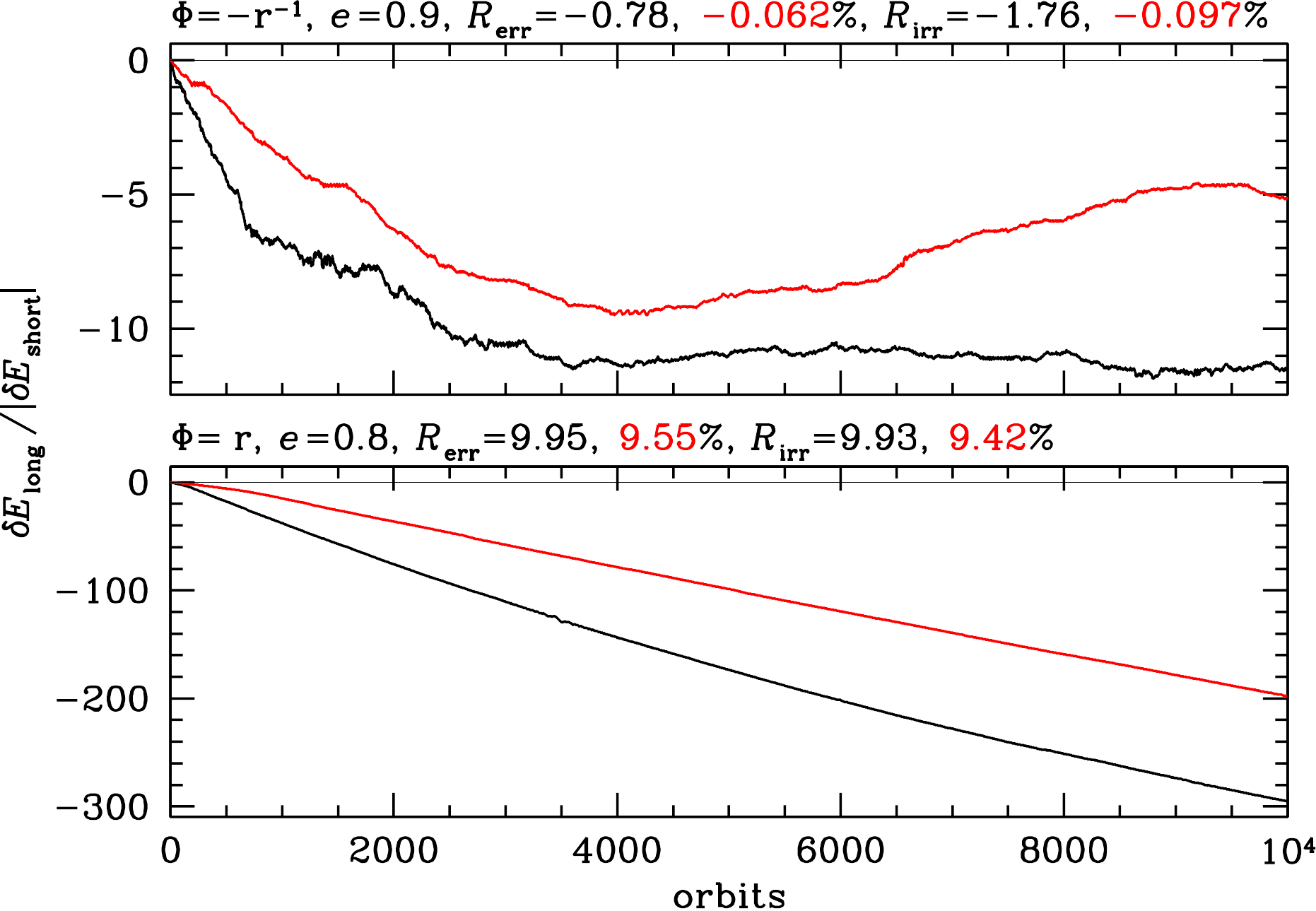}\hfill\\[1ex]
	
	\hfill\includegraphics[width=84mm]{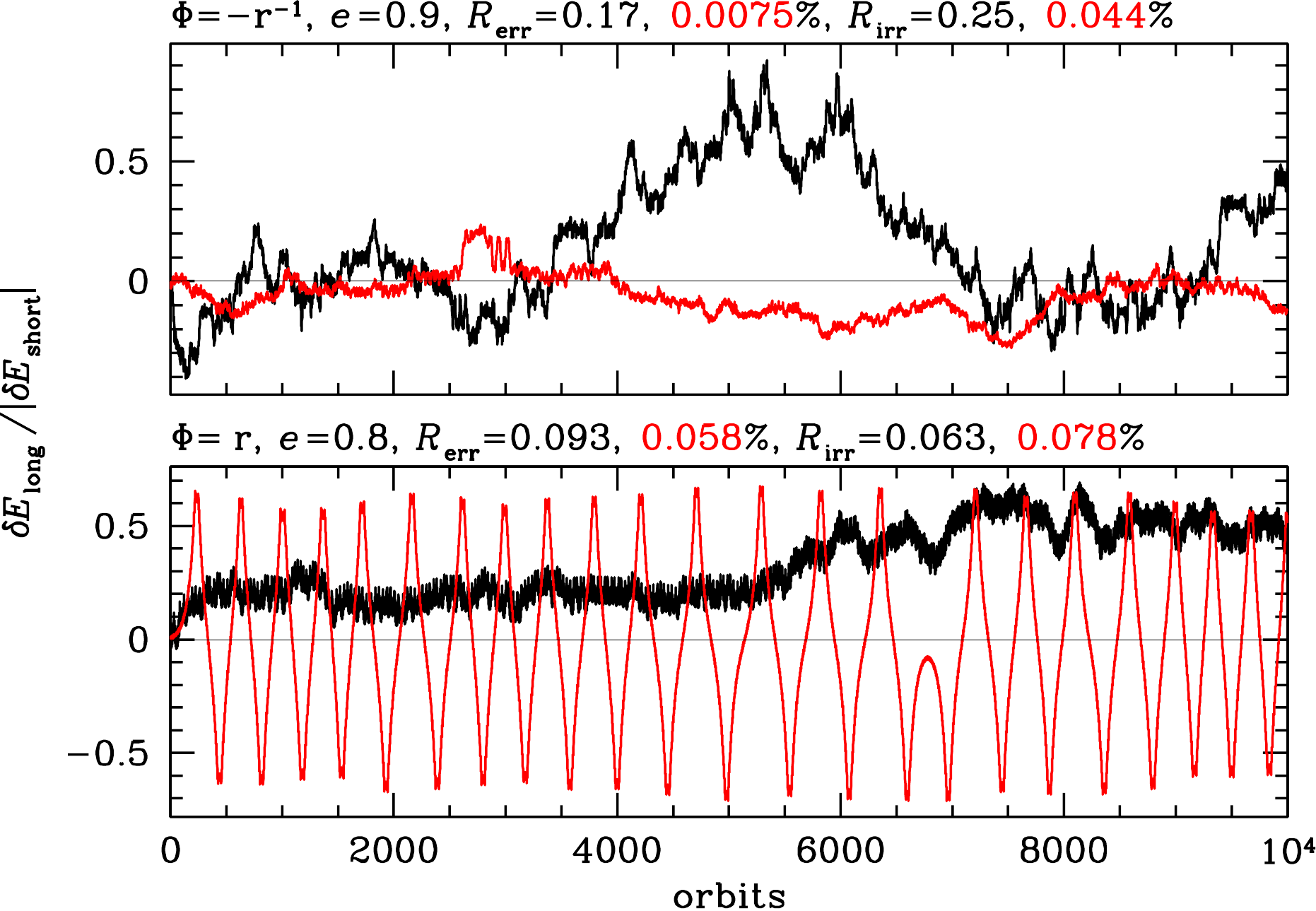}\hfill
	\caption{\label{fig:symm:dE:mid}
	As Fig.~\ref{fig:symm:dE:apo} but with the integration started half way between apo- and peri-centre. The top panels are for initial $\tau_{1/2}=T_0$, while the bottom panels are for initial $\tau_{1/2}=\sqrt{T_0T_1}$; in both cases $h_{1/2}=h_{\mathrm{block}}(T_0)$.
	}
\end{figure}
%%%%%%%%%%%%%%%%%%%%%%%%%%%
\subsection{\changed{Reversible adaptation of a continuous time step variable}}
\label{sec:symm}
\changed{
A direct porting of equation~\eqref{eq:explicit:mu} to discrete step sizes obtains a scheme only slightly better than the forward scheme, as described in the appendix. The problem is that for certain intermediate values of $T$ the discrete step size $h$ flips between two values bracketing $T$. These flips are reversible only if step-size changes are unlimited, which allows the possibility of arbitrary long steps.

To avoid too long steps, one must not allow this flipping of $h$. This can be achieved by following} the initial scheme~\eqref{eq:explicit:mu} with an auxiliary \changed{continuous} variable $\tau$ \changed{instead of $h$} and set
\begin{equation} \label{eq:h:tau}
	h_{n+1/2} = h_{\mathrm{block}}(\tau_{n+1/2}).
\end{equation}
However, implementing this directly via 
\begin{equation} \label{eq:tau:fail}
	\tau_{n+1/2}=T_n^2/\tau_{n-1/2}
\end{equation}
(for $\mu$ the geometric mean) fails. The problem is that $\tau$ oscillates around $T$ with amplitude increasing at each change in step size\footnote{
	This increasing oscillation amplitude seems at odds with the reversible nature of the method. However, this can be understood in analogy to linear instability where \changed{time symmetry} implies that each growing mode is accompanied by a decaying mode of no practical significance.}. If $h$ remains unchanged, the relation~\eqref{eq:tau:fail} places $\ln(\tau_{n-1/2}),\;\ln(T_n)$ and $\ln(\tau_{n+1/2})$ on a line when plotted against time \changed{(rather than step index)}. So, to avoid the amplification of oscillations when $h$ changes, \changed{one may} simply retain that linear relation, which gives
\begin{equation} \label{eq:symm:flip}
	\tau_{n+1/2}=T_n\left(\frac{T_n}{\tau_{n-1/2}}\right)^{\frac{h_{n+1/2}}{h_{n-1/2}}}
\end{equation}
instead of~\eqref{eq:tau:fail}. Together with equation~\eqref{eq:h:tau}, this is an implicit relation for $h_{n+1/2}$ and $\tau_{n+1/2}$ given the previous values and $T_n$. Again, solving these relations is straightforward, obtaining the scheme
\begin{equation} \label{eq:symm}
	\delta r = \left\{\begin{array}{lllcll}
	-1 \;\;& \text{if} \;\; & h_{n-1/2} \tau_{n-1/2}^2 & \le & \tfrac{1}{2} T_n^3 & 
	\;\; \text{and block-step allows\hspace*{-5mm}}, \\[0.5ex]
	\phm0 & \text{if} & h_{n-1/2} \tau_{n-1/2} & \le & T_n^2,\\[0.5ex]
	+1 & \multicolumn{4}{l}{\text{otherwise.}}
	\end{array}\right.
\end{equation}
The resulting \changed{ratio of long- to short-}term energy error for the two test orbits is shown in Fig.~\ref{fig:symm:dE:apo} when started at apo-centre and Fig.~\ref{fig:symm:dE:mid} when started half way between apo- and peri-centre. In the latter case, one sees a significant difference between an orbit integration started with $\tau_{1/2}=T_0$ or $\tau_{1/2}=\sqrt{T_0T_1}$ (and $h_{1/2}=h_{\mathrm{block}}(\tau_{1/2})$ in both cases). This difference can be understood by a significant number of irreversible steps (as reported via $R_{\mathrm{irr}}$).

\begin{figure}
	\begin{center}
		\includegraphics[width=75mm]{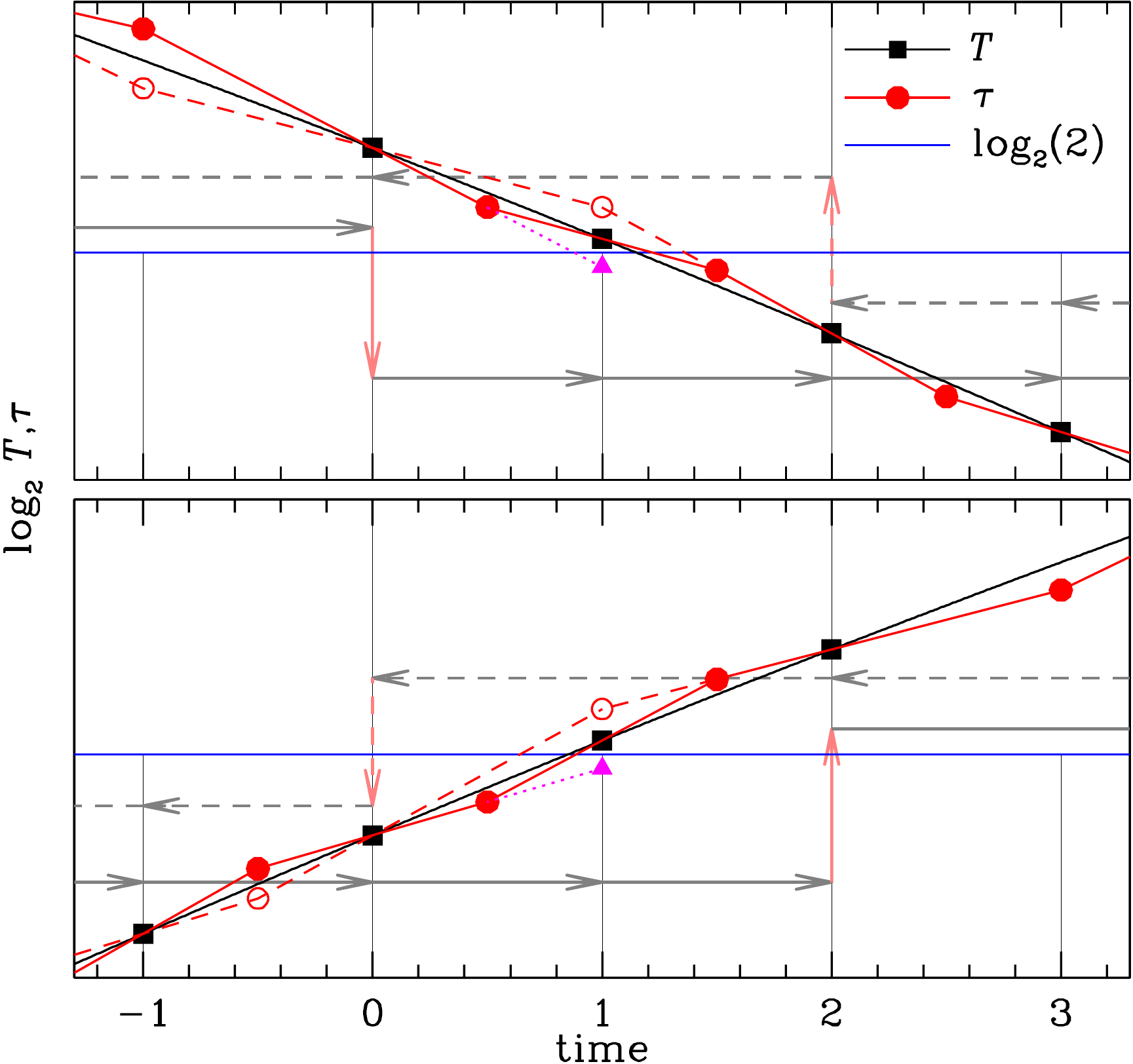}
	\end{center}
	\caption{\label{fig:symm:irreversible}
	\changed{Mechanics} of irreversible step-size changes with scheme~\eqref{eq:symm}: two small steps with $h=1$ \changed{(solid arrows)} are taken from $t=\changed{0}$ to $t=\changed{2}$, but in reversed time a single step with $h=2$ is chosen \changed{(dashed arrow)}. The auxiliary $\tau$ (circles) in the middle of each step is reversibly evolved (red lines) via equation~\eqref{eq:symm:flip} using the time-step function $T$ evaluated at the end of each step (squares). The \changed{blue} horizontal line indicates \changed{a value of 2: if $\tau$ predicted for the middle of a long step is above this line, a long step is taken, otherwise two short steps.} At time $t=\changed{0}$ in the forward direction, $h=2$ would result in $\tau<2$ \changed{(triangle) predicted for $t=\changed{1}$,} forcing \changed{two steps} with $h=1$ to $t=\changed{2}$. However, at $t=\changed{2}$ in the backward direction, $h=2$ \changed{is acceptable}, resulting in a different $\tau$ sequence (open circles \changed{\& dashed lines}). \changed{This mechanism can occur for $\dot{T}<0$ (top) or $\dot{T}>0$ (bottom).}
	}
\end{figure}

Fig.~\ref{fig:symm:irreversible} shows the mechanics of irreversible step-size changes with this scheme. \changed{They an occur for both signs of $\dot{T}$ and originate from an integration/extrapolation error of $\tau$, which causes the value for $\tau$ in the middle of a long step to be different in the backward direction, if that step was not taken.} The chance for such \changed{integration errors} and hence irreversible step-size changes is greatly enhanced if the auxiliary quantity $\tau$ oscillates around the time-step function \changed{(as it does to some extent in Fig.~\ref{fig:symm:irreversible})}. If $\tau_{n-1/2}=T_{n-1/2}$, relation~\eqref{eq:symm:flip} corresponds to a linear extrapolation of $\ln(\tau)=\ln(T)$ in time. But if $\tau_{n-1/2}\neq T_{n-1/2}$ or if $\ln(T)$ has curvature, this extrapolation goes slightly wrong and causes step-to-step oscillations of $\tau$ around $T$, like those \changed{detected} by \citeauthor{CirilliHairerLeimkuhler1999} for \changed{non-discrete} $h$. This explains why carefully choosing the initial $\tau$ greatly reduces the number of irreversible steps (and with it the energy error) \changed{reported} in the bottom panels of Fig.~\ref{fig:symm:dE:mid}.

%%%%%%%
%%%%%%%

%%%%%%%%%%%%%%%%%%%%
\subsection{Can the situation be improved?}
\label{sec:symm:suppress}

However, this reduction of irreversible steps through careful choice of the initial $\tau$ works only as long as the time-step function is sufficiently smooth. This is not necessarily the case in $N$-body simulations, where $T(\xi)$ may fluctuate. 

\changed{
In order to reduce the number of irreversible step-size changes, one may devise techniques to suppress these oscillations, even though this inevitably introduces another type of irreversibility. One technique is to re-align $\tau=T$ when $\log_2\tau$ is close to a half-integer, i.e. as far away from any step-size changes as possible. The resulting test orbit integrations give comparable results to those obtained with a careful choice of the first step size (Fig.~\ref{fig:symm:dE:mid}, bottom panels).
}

%%%%%%%%%%%%%%%%%%%%%%%%%%%%%%%%%%%%
\changed{Another idea is to} avoid irreversible step-size changes by means of the try-and-reject method of section~\ref{sec:care} (which violates time reversibility in a more subtle and less harmful way than irreversible steps).
Because one cannot detect the violation of reversibility in time to avoid it, the only possibility is to try avoiding stepping errors. Then, instead to prefer the longer step in case of ambiguity (as per the argument of section~\ref{sec:prob:reverse}), one tries the shorter step but rejects it if the longer step is found acceptable. \changed{However, this introduces a new type of irreversible rung change, because} not all stepping errors of the scheme~\eqref{eq:symm} are irreversible.

%%%%%%%%%%%%%%%%%%%%%%%%%%%%%%%%%%%%%%%%%%%%%%%%%%%%%%%%%%%%%%%%%%%%%%%%%%%%%%%%
\section{Integrating the step size}
\label{sec:diff}
The time stepping schemes considered in the previous two sections attempt to synchronise the right- and left-hand sides of $h=T$ (equation~\ref{eq:problem}), respectively. I now consider schemes which instead synchronise a differential form, like 
\begin{subequations}
\begin{eqnarray}
	\Delta h^{-1}&=&\Delta T^{-1}
	\qquad\text{or}\\
	\Delta \ln(h)&=&\Delta \ln(T).
\end{eqnarray}
\end{subequations}
The left-hand and right-hand sides of these relations can be expressed as finite difference and differential, respectively. For \changed{non-discrete} step sizes, this obtains the schemes
\begin{subequations}
	\label{eq:h:diff}
\begin{equation}
	\label{eq:h:diff:harm}
	\frac{1}{h_{n+1/2}} - \frac{1}{h_{n-1/2}}
%	= T_n \tdiff{T_n^{-1}}{t}
	= - \frac{\dot{T}_n}{T_n}
\end{equation}
\citep{HairerSoderlind2005} or
\begin{equation}
	\label{eq:h:diff:geom}
	\ln h_{n+1/2} - \ln h_{n-1/2}
%	= T_n \tdiff{\ln(T_n)}{t} 
	= \dot{T}_n.
\end{equation}
\end{subequations}
According to \citeauthor{HairerSoderlind2005} and my own experiments, these schemes perform very well for continuous step sizes (no block-step)\changed{: even for chaotic orbits $h$ closely follows $T$}. Most importantly, this scheme avoids step-by-step oscillations of $h$, and hence is unlikely to suffer from any significant number of irreversible step-size changes, when adapted to the block-step.

The price to pay is the need to compute the time derivative $\dot{T}$ of the time-step function. This is usually not efficiently possible in large-$N$ simulations, but is a reasonable option for small-$N$ methods.

%%%%%%%
\begin{figure}
	\hfill\includegraphics[width=84mm]{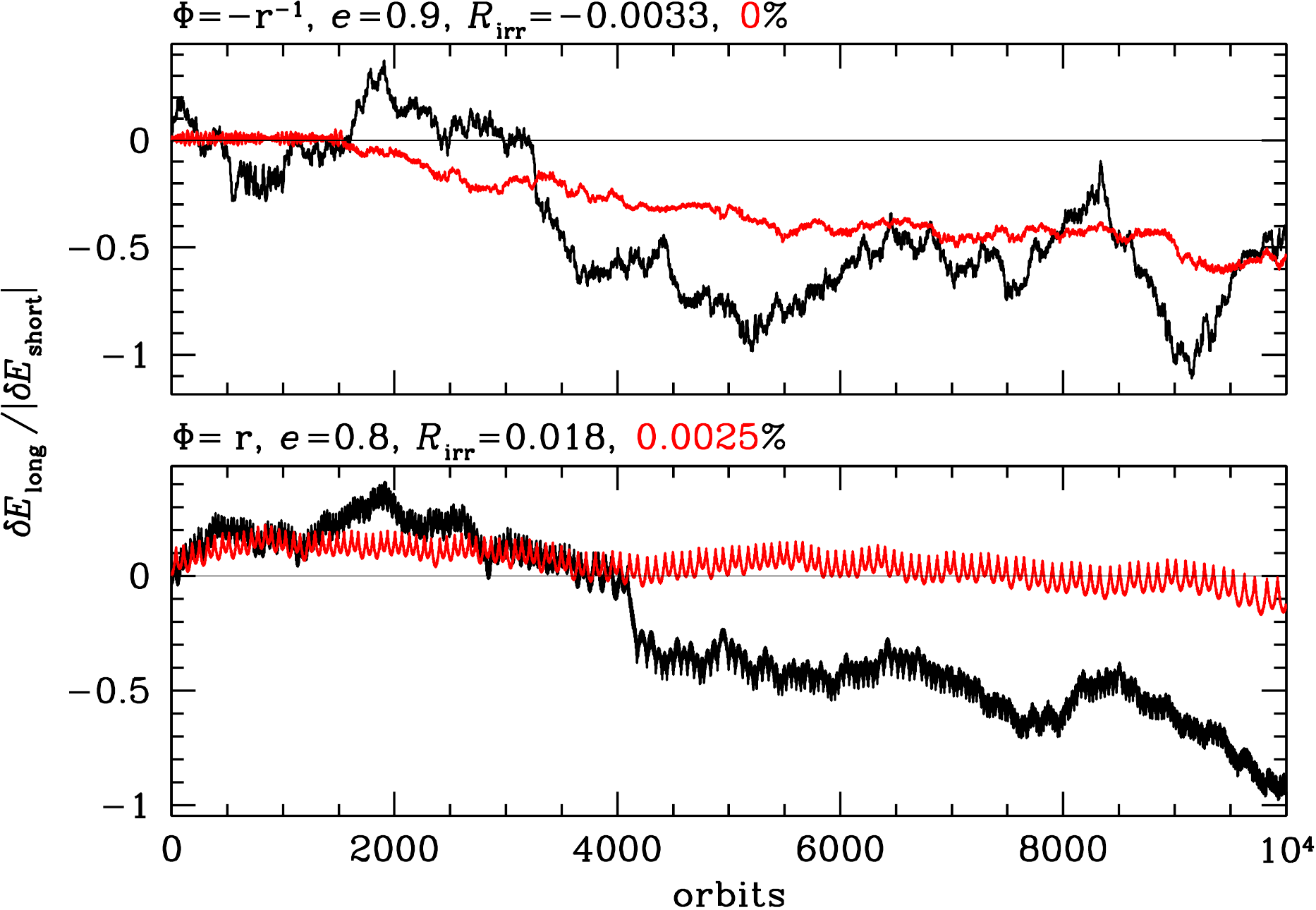}\hfill
	\caption{\label{fig:intg:dE}
	As Fig.~\ref{fig:fwrd:dE} but using scheme~\eqref{eq:diff:geom}. \changed{For $\eta=0.02$,} there are only \changed{8} and \changed{11} irreversible steps for the Kepler (top) and cusp (bottom) orbit, respectively\changed{, while for $\eta=0.01$ the numbers are 0 and 1}. See also Fig.~\ref{fig:intg:kepler:dE} for an integration of the $e=0.9$ Kepler orbit over $10^6$ periods. 
	}
\end{figure}
\begin{figure}
	\begin{center}
		\includegraphics[width=75mm]{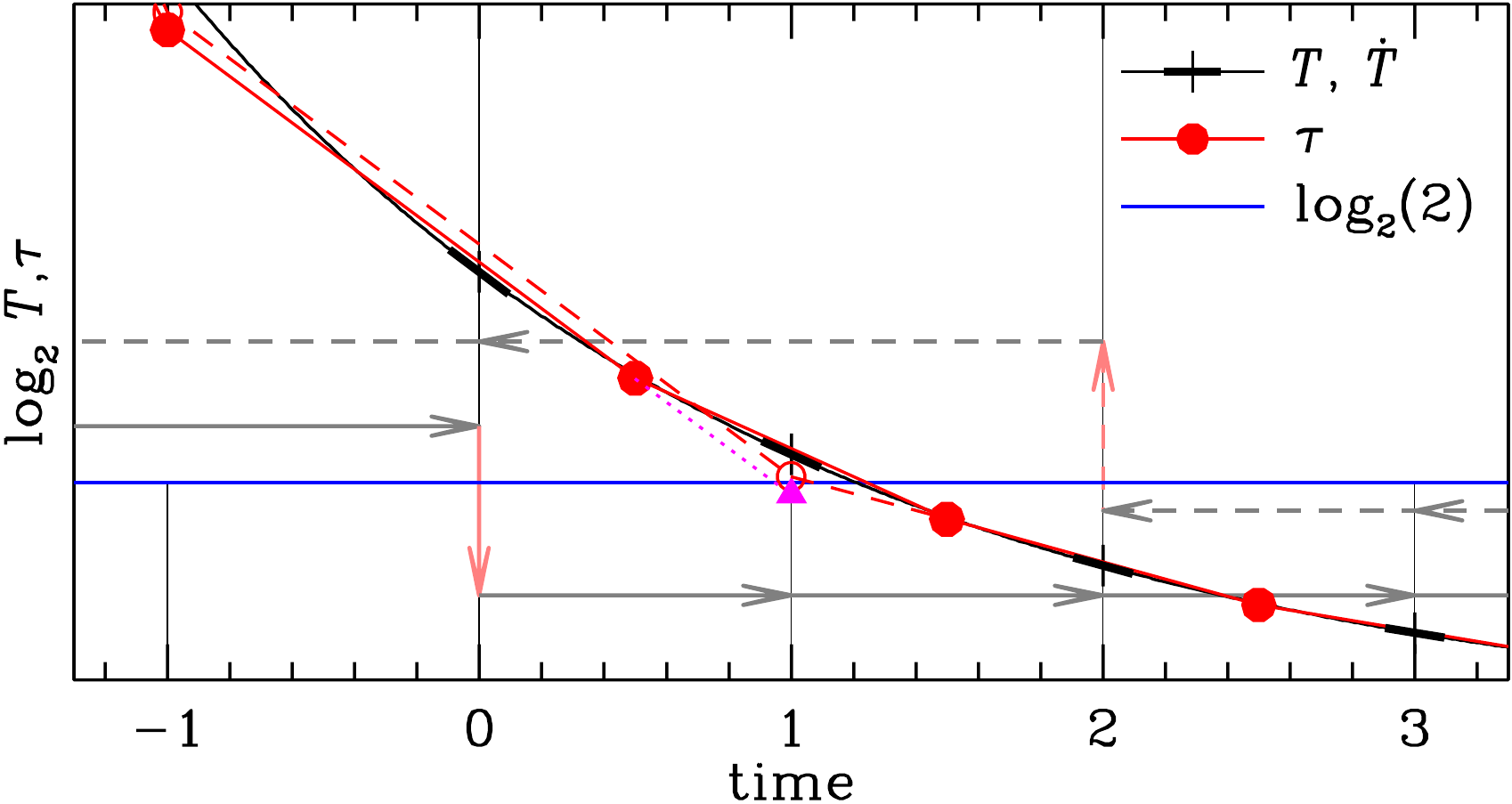}
	\end{center}
	\vspace*{-3mm}
	\caption[]{\label{fig:intg:irreversible}
	\changed{The mechanics} of an irreversible step-size change with the scheme~\eqref{eq:diff:geom} \changed{at $\dot{T}<0$ are analogous to that of scheme~\eqref{eq:symm} (see top panel of Fig.~\ref{fig:symm:irreversible}), except that} $\tau$ (circles) in the middle of each step is reversibly evolved \changed{more accurately}, via equation~\eqref{eq:diff:tau:geom} using \changed{the time step function $T$ and its derivative $\dot{T}$} evaluated at the end of each step. \changed{Again, the irreversibility occurs because $\tau(1)$ predicted in the forward (triangle) and backward (open circle) directions differ and between them bracket the critical threshold for switching the discrete step size $h$. This can also occur for $\dot{T}>0$, analogously to the bottom panel of Fig.~\ref{fig:symm:irreversible}.}
	}
\end{figure}

%%%%%%%
Since the block-step does not allow incremental changes to $h$, one cannot use equations~\eqref{eq:h:diff} directly, but must use an auxiliary variable $\tau$. One must then also account for the discrete step over which $\tau$ is integrated, giving
\begin{subequations} \label{eq:diff:tau}
	\begin{eqnarray}
	\label{eq:diff:tau:harm}
	\frac{1}{\tau_{n+1/2}} &\changecolor=&
	\frac{1}{\tau_{n-1/2}}
	 - 	\frac{h_{n-1/2}+h_{n+1/2}}{2} \frac{\dot{T}_n}{T_n^2},
	\\ \label{eq:diff:tau:geom}
	\ln \tau_{n+1/2} &\changecolor=&
	\ln \tau_{n-1/2}
	 \,{\changecolor +}\,\frac{h_{n-1/2}+h_{n+1/2}}{2} \frac{\dot{T}_n}{T_n}
	\end{eqnarray}
in place of equations~\eqref{eq:h:diff}.
\end{subequations}
Combined with equation~\eqref{eq:h:tau}, one again has trivially solvable implicit relations for $h_{n+1/2}$. When restricting rung changes to $|\delta r|\le1$ as usual, this obtains the stepping schemes
\begin{subequations} \label{eq:diff}
\begin{eqnarray} \label{eq:diff:harm}
	\delta r &=& \left\{\begin{array}{lllcll}
	-1 \;\;& \text{if} \;\; & \displaystyle \frac{2}{\tau_{n-1/2}} -\frac{3d}{T_n} & \le & \displaystyle \frac{1}{h_{n-1/2}} & \;\; \text{and block-step allows}, \\[2.5ex]
	\phm0 \;\;& \text{if} \;\; & \displaystyle \frac{1}{\tau_{n-1/2}} -\frac{d}{T_n} & \le & \displaystyle \frac{1}{h_{n-1/2}}
	\\[2.5ex] 
	+1 & \multicolumn{4}{l}{\text{otherwise,}}
	\end{array}\right. \hspace*{-5mm}
	\\[1ex]
	\label{eq:diff:geom}
	\delta r &=& \left\{\begin{array}{lllcll}
	-1\;\;&\text{if} \;\; & \rho_{n-1/2}-\frac{3}{2}(d/\ln2) & \le & r_{n-1/2}-1 &
	\;\; \text{and block-step allows}, \\[1ex]
	\phantom{-}0 \;\;&\text{if} \;\; & \rho_{n-1/2}-\phantom{\frac{3}{2}}(d/\ln2) & \le & r_{n-1/2},\\[1ex]
	+1\;\;&\multicolumn{4}{l}{\text{otherwise.}}
	\end{array}\right. \hspace*{-5mm}
\end{eqnarray}
\end{subequations}
where $d \equiv h_{n-1/2}(\dot{T}_n/{T_n})$, while $\rho\equiv-\log_2(\tau/h_{\max})$ is a continuous rung variable replacing $\tau$. To initialise $\tau$, one must integrate the first half time step, i.e.
\begin{subequations} \label{eq:diff:init}
\begin{eqnarray}
	\label{eq:diff:harm:init}
	\frac{1}{\tau_{1/2}} &=& \frac{1}{T_0} - \frac{h_{1/2}\dot{T}_0}{2 T_0^2},
	\\
	\label{eq:diff:geom:init}
	\tau_{1/2} &=& T_0 \exp\left(\frac{h_{1/2}\dot{T}_0}{2T_0}\right).
\end{eqnarray}
\end{subequations}
Fig.~\ref{fig:intg:dE} shows the run of the \changed{ratio of long- to short-term} energy error \changed{for} this scheme \changed{and the two} test orbits. There are still very few irreversible step-size changes\changed{, which can occur for either sign of $\dot{T}$. These are caused by the same basic mechanism that is also responsible for irreversible step-size changes with the scheme~\eqref{eq:symm}, see Fig.~\ref{fig:intg:irreversible}. However, because the time integration of $\tau$ is more accurate and avoids oscillations,} the schemes~\eqref{eq:diff} only suffer very occasionally from  irreversible rung changes. \changed{For the $e=0.9$ Kepler orbit, for example, no irreversible step-size changed occurred for $\eta=0.01$ over $10^4$ orbits (the small-amplitude trend of the long-term energy for this orbit must be owed to other sources, see also the last paragraph of Section~\ref{sec:xtra}).}

One can differentiate the relation between $h$ and $T$ once more and numerically integrate a second-order differential equation for $\tau$, for example using the kick-drift-kick method. This obtains $\tau_{n+1}$  reversibly predicted from $\tau_{n}$ and $\dot{\tau}_n$ and enables the time-step condition $h_{n+1/2}\le\mu(\tau_n,\tau_{n+1})$. Such an approach is conceptually very similar to a zonal time-step function $T=T(\vec{x})$, discussed in section~\ref{sec:zonal} below. It is quiestionable, however, whether this significantly reduces the chances of irreversible step-size changes, since the integration schemes~\eqref{eq:diff:tau} are already second-order accurate.

%%%%%%%
%%%%%%%

%%%%%%%%%%%%%%%%%%%%%%%%%%%%%%%%%%%%
\section{Discussion}
\label{sec:discuss}
The usage of individual adaptive time-step sizes in $N$-body simulations can result in enormous efficiency savings. Without these, hardly any of the many computer simulations of stellar dynamics, large-scale structure formation, and galaxy formation would have been possible. However, this technique still suffers from fundamental limitations in that almost all contemporary implementations violate time \changed{symmetry} at the basic level, by setting the step size equal to its desired value at the \changed{beginning of each step} (the forward method). Such violations are well known to affect the long-term stability of the simulations, resulting in artificial dissipative behaviour, which is often (but not necessarily) revealed by a secular drift of the total energy.

%%%%%%%%%%%%%%%%%%%%%%%%%%%%%%%%%%%%
\subsection{Can exact reversibility be achieved?}
\label{sec:reversible?}
Adapting the size $h$ of each time step reversibly is non-trivial because the 
time-step function $T(\xi)$, which provides the ideal value for $h$ given the state $\xi$ of the trajectory, can only be evaluated at the start of each step to inform the choice of its size $h$. Thus, $T$ and $h$ are not naturally synchronised and to achieve time reversibility some form of synchronisation is necessary.

This problem of synchronising $h$ and $T$ also exists in the case of a single continuous step size (unconstrained by the block-step scheme). This simpler situation has been well studied and several solutions have been proposed and demonstrated to give good results \citep[see][for an overview]{HairerLubichWanner2002}. These methods essentially \emph{integrate} the step size itself in a reversible way.

However, porting these methods to the block-step scheme and at the same time retaining reversibility appears impossible (at least when efficiency is retained. So how does the block step render the problem so much more difficult?

%%%%%%%
\begin{figure}
	\hfill\includegraphics[width=60mm]{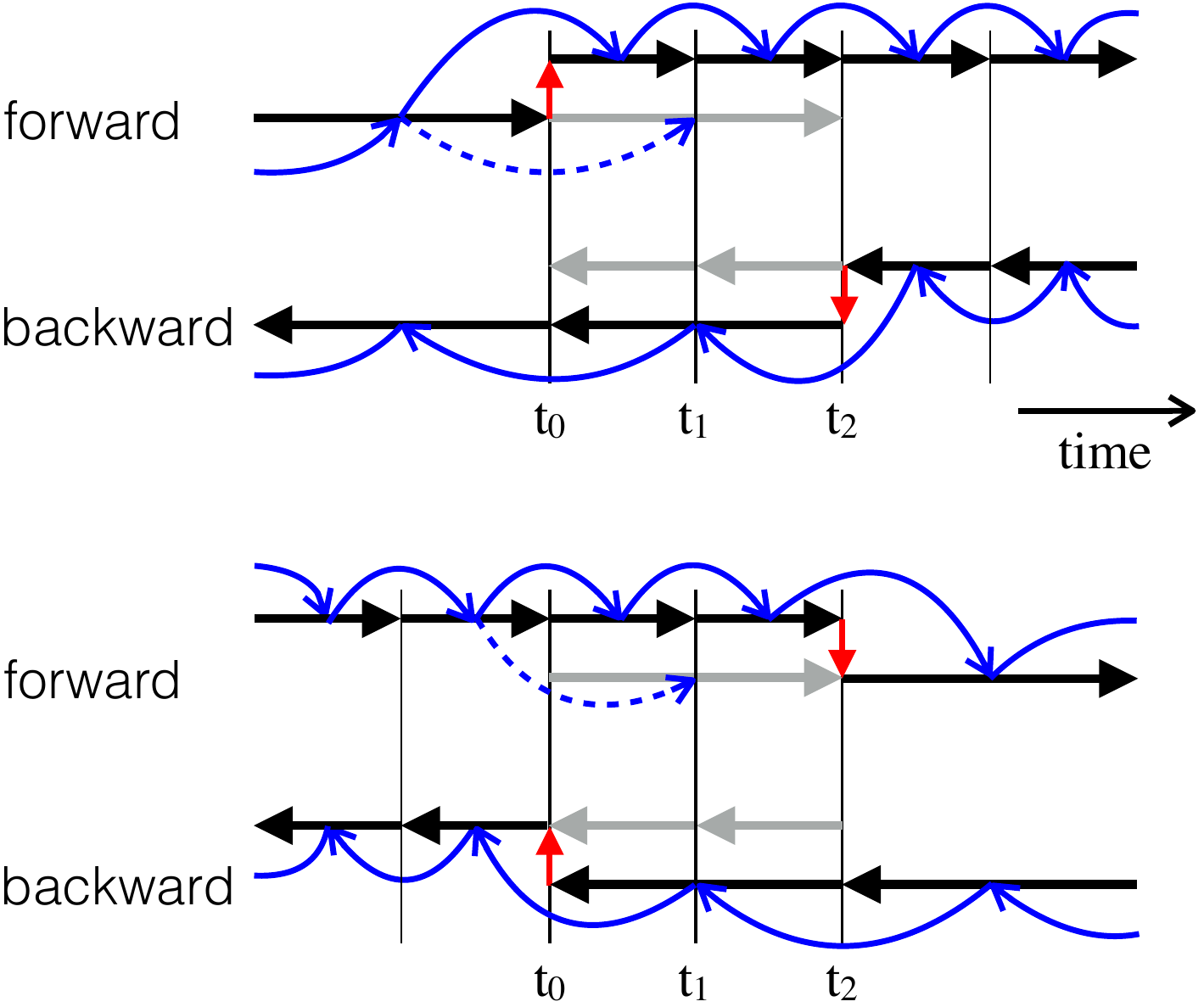}\hspace{10mm}\hfill
	\caption{\label{fig:irr:horizon}
	Schematics of irreversible step-size change\changed{s, see also Figs.~\ref{fig:symm:irreversible} \& \ref{fig:intg:irreversible}}. Time runs horizontally and step size vertically, \changed{steps and step-size changes are shown as horizontal and vertical arrows, respectively, and the reversible integration of $\tau$ is indicated by blue arrows. \textbf{Top}:} in the forward direction at time $t_0$, maintaining the long step size (grey) \changed{would} not meet the time-stepping conditions reversibly predicted (\changed{dashed}) for $t_1$. Therefore, the step size must be changed and two shorter steps are made to time $t_2$. However, in the backward direction at $t_2$, the prediction for the time-stepping condition at $t_1$ allows a long step to be taken instead of two short ones (grey). \changed{\textbf{Bottom:} The mechanism can also occur when the step size is increased.}
	}
\end{figure}
%%%%%%%
%%%%%%%%%%%%%%%%%%%%%%%%%%%%%%%%%%%%
\subsubsection{\changed{Discreteness and the} horizon of predictability}
\label{sec:horizon}
The effect of the block-step scheme is to discretise the step sizes $h$. Without such discretisation synchronisation between $T$ and $h$ can be achieved by \emph{exactly} solving an explicit reversible adaption condition, such as equations~\eqref{eq:explicit:mu} or~\eqref{eq:h:diff}. When discretising $h$, this approach is no longer viable, as none of the discrete values will solve the reversibility condition exactly.

Instead, \changed{one has to follow an explicit and reversible adaption} condition with an auxiliary \changed{continuous} variable $\tau$\changed{, which} in turn is then used to inform the choice for the discretised $h$. Complications arise because the time evolution of $\tau$ depends on the actual discrete step sizes $h$ \changed{used}, which creates a mutual dependency between $h$ and $\tau$ and hence an implicit relation for $h$. Because only a finite number of discrete values are allowed for $h$, this implicit relation can be solved without iterations\changed{, using the longest allowed step in case of ambiguity.

The reason for preferring the longest allowed step is that the merit of the alternative, a pair of two shorter steps, cannot be assessed without violating time symmetry, since the second short step is still in the future and beyond the horizon of predictability.

The integration of the continuous step-size variable $\tau$ is inevitably subject to errors and, as a consequence, different values are obtained by different sequences of steps. Consider the situation depicted in Fig~\ref{fig:irr:horizon}.} If two short steps are taken between $t_0$ and $t_2$. then the value $\tau_{\mathrm{for}}$ for $\tau(t_1)$ predicted \changed{at time $t_0$} in the forward direction differs \changed{(in general)} from $\tau_{\mathrm{back}}$, the value predicted for the same quantity at time $t_2$ in the backward direction.

\changed{If this difference is that between taking the long step from $t_0$ to $t_2$ or two shorter steps, an irreversible step-size change results. Thus,} if two short steps are taken instead of one long one, then the value $\tau_{\mathrm{back}}$ for the alternative longer step is beyond the horizon of predictability at the moment $\tau_{\mathrm{for}}$ is computed. \changed{This mechanism only produces irreversibly step-size changes with $h_{\mathrm{for}} < h_{\mathrm{back}}$ but of either sign of $\dot{T}$. This is exactly what I find for the schemes~\eqref{eq:symm} and \eqref{eq:diff}: all irreversible step-size changes are of this kind and their frequency is substantially larger} when removing the block-step's hierarchical synchronisation (but still requiring discretised step sizes). 

%%%%%%%
\begin{figure}
	\hfill\includegraphics[width=70mm]{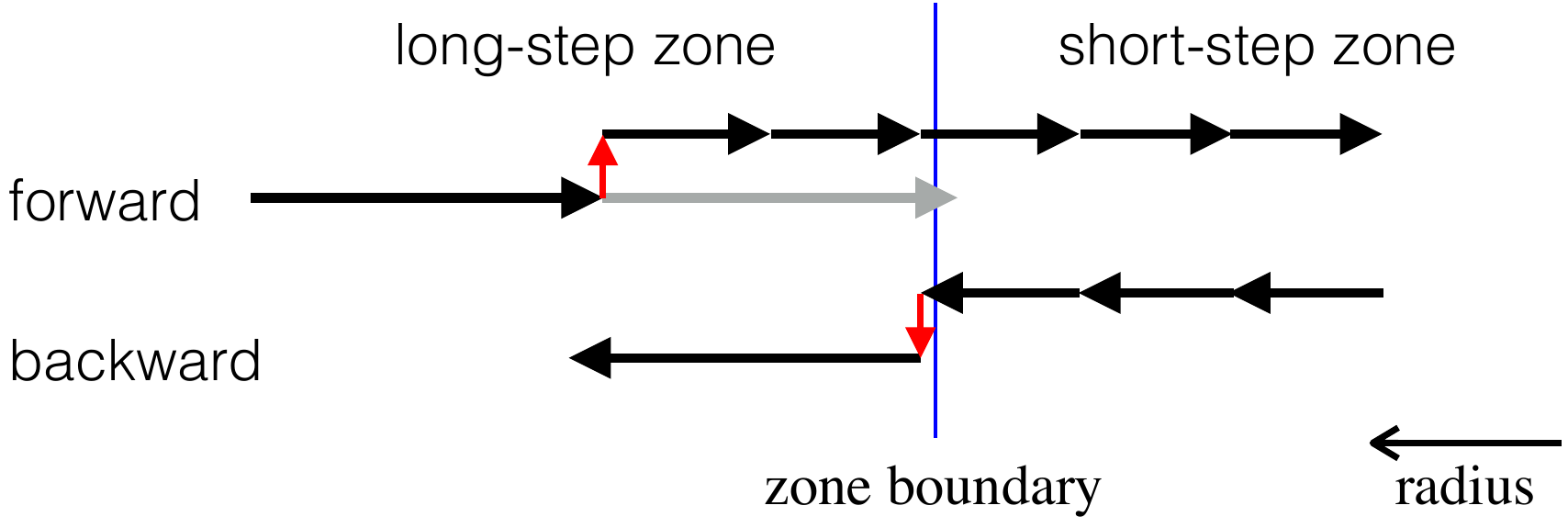}\hfill
	\caption{\label{fig:irr:zonal}
	Schematics of an irreversible step-size change with a zonal time-step function. Radius runs horizontally and step size vertically. An inward moving particle encounters a zone boundary such that another long step (grey) would move it into the short-step zone. Hence, two shorter steps are made instead. If these do not penetrate the short-step zone, a different sequence of steps will be used in the backward direction.
	}
\end{figure}
%%%%%%%
%%%%%%%%%%%%%%%%%%%%%%%%%%%%%%%%%%%%
\subsubsection{The case of zonal time-step functions}
\label{sec:zonal}
A zonal time-step function $T=T(\vec{x})$ depends only on the particle position. This is useful in simulations of near-equilibrium galactic dynamics, when the local orbital time is well described by a simple function of radius. The discrete step sizes allowed by the block-step then correspond to radial zones \citep{Sellwood2014}.

For zonal time-step functions (and using the kick-drift-kick leapfrog) it seems that exact reversibility can be obtained with the block-step, because the value $T_{n+1}$ can be readily computed from the information known at time $t_n$.
\citeauthor{Sellwood2014}, for example, uses for $h_{n+1/2}$ the minimum of the discrete step sizes required at $t_n$ and $t_{n+1}$. However, the trajectory is subject to integration errors and hence so is the step size, such that the problem discussed in the previous section still pertains, as explained in Fig.~\ref{fig:irr:zonal}, and even for zonal time-step functions exact reversibility is elusive.

%%%%%%%%%%%%%%%%%%%%%%%%%%%%%%%%%%
\subsection{\boldmath \changed{Applicability to} contemporary $N$-body methods}
The importance of the adverse effects of irreversible step-size adaptation varies \changed{considerably} between different applications of the $N$-body method \changed{and so does the necessity to and potential benefit of applying any of the methods considered in this study. One potentially serious problem is the noise level, as quantified by $|\dot{T}|$, of any practical time-step-function implementation, which for most of the adaptation schemes will result in poorer reversibility and hence long-term stability than for single-orbit integrations.}

%%%%%%%%%%%%%%%%%
\subsubsection{Simulations of collision-less dynamics}
In contemporary $N$-body simulations of collision-less stellar dynamics (galaxy \changed{formation and} interactions\changed{, as well as} large-scale structure \changed{formation}) the adverse effects \changed{of irreversible step-size adaptation from the naive forward scheme} are presumably tolerable, i.e.\ the resulting errors remain small over the duration of the simulations (though this has not been rigorously validated), for two reasons. First, such simulations only cover $\sim10^{2-3}$ dynamical times, many fewer than the test orbit integrations presented here.

Second, the time-step function most commonly used in such simulations, $T=\eta\sqrt{\epsilon/|\vec{a}|}$, where $\epsilon$ is the softening length and $\vec{a}$ the acceleration, varies only weakly over typical orbits. (In fact, this time-step function is sub-optimal, since for the vast majority of orbits it gives unnecessarily \changed{short} time steps, see \citealt{ZempEtAl2007} for a better alternative.) \changed{Moreover, in situations where $\epsilon$ is smaller than the inter-particle separation close encounters will result in $|\dot{T}|\not{\ll}1$. For most of the stepping schemes discussed, this inevitably leads to a degradation of the performance compared to the simple test-orbit integrations considered in this study. Thus, if a scheme with considerably \changed{better reversibility} than the state-of-the-art forward method is required, some form of adaptive force softening appears desirable.}

%%%%%%%%%%%%%%%%%
\subsubsection{Simulations of collisional dynamics}
Simulations of star clusters, on the other hand, span usually $\sim10^{4-6}$ dynamical times, when adverse effects from violating time symmetry may well become important. However, such simulations are usually done very carefully, typically by controlling or monitoring the total energy error and hence guarding against artificial secular dissipation. 

In this case, the total net energy error may still be dominated by the adverse effects of the step-size adaptation. If this is so, then a more careful scheme for adapting the step sizes may improve the efficiency of collisional $N$-body simulations. \changed{The problem of fluctuating time-step functions is likely much less of a problem than for collision-less dynamics, simply because shorter time steps are chosen for enhanced accuracy. This directly reduces $|\dot{T}|$ and with it the rate of irreversible step changes for all of the adaptation schemes considered in this study and in contrast to the forward method.}

%%%%%%%
\begin{figure}
	\hfill\includegraphics[width=84mm]{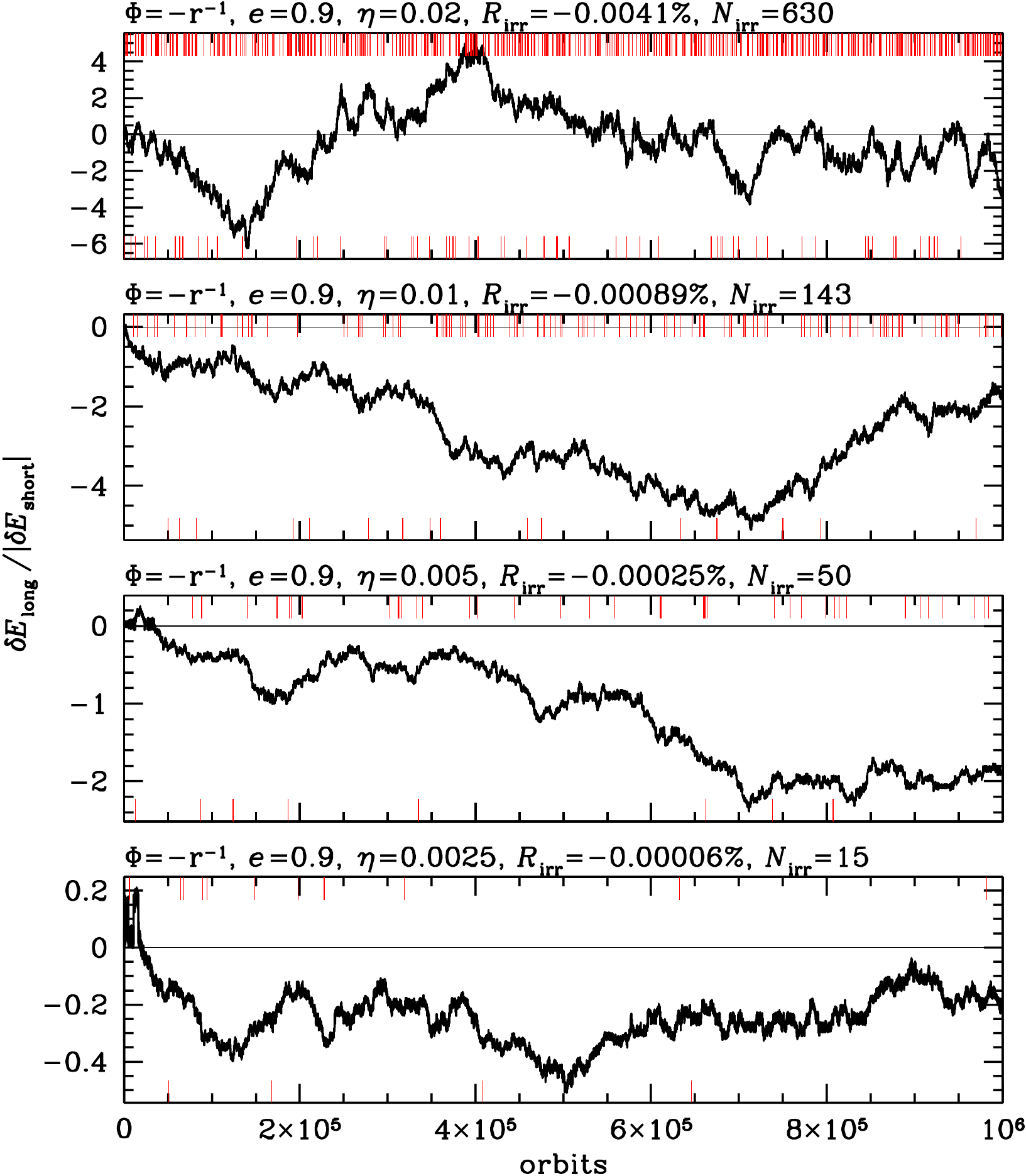}\hfill
	\caption{\label{fig:intg:kepler:dE}
	As Fig.~\ref{fig:intg:dE} but integrating 100 times longer and only for the Kepler orbit using four decreasing values of $\eta$ as indicated (note the different vertical scales). \changed{Thin red markers on the top and bottom of each panel indicate irreversible step-size changes for $\dot{T}<0$ and $\dot{T}>0$, respectively, ($h<h_{\mathrm{back}}$ in each case), while $N_{\mathrm{irr}}$ is their total number.}
	}
\end{figure}
%%%%%%%
%%%%%%%%%%%%%%%%%
\subsubsection{Simulations of planetary systems}
Finally, $N$-body simulations of planetary systems span $\sim10^{8-12}$ dynamical times, enough for even mild secular dissipation to accumulate. In order to avoid this, such integrations are typically done with a single global time-step size (which may be adapted).

Of all the block-step based adaptation schemes considered in this study, only those of Section~\ref{sec:diff} appear at all suitable for integrations of planetary systems. These techniques avoid oscillations of the auxiliary continuous step-size variable $\tau$ and only suffer from the apparently unavoidable irreversibilities of the type described in Figs.~\ref{fig:intg:irreversible}\&\ref{fig:irr:horizon} and Section~\ref{sec:horizon}.

In order to assess how the rate of these irreversible step-size changes depends on the intended integration accuracy, i.e.\ on the parameter $\eta$ of the time-step function~\eqref{eq:T:omega}, I ran some integrations of the $e=0.9$ Kepler orbit for $10^6$ orbits \changed{(100 times longer than in Fig.~\ref{fig:intg:dE})}, see Fig.~\ref{fig:intg:kepler:dE}. Encouragingly, the net-rate $R_{\mathrm{irr}}$ of irreversible step-size changes (but also their total number) decrease like $\eta^2$ (the same scaling as for the energy error). This is because decreasing $\eta$ increases the integration accuracy not only of the trajectory, but also of the continuous step-size variable $\tau$, and hence reduces the chances of \changed{irreversible step-size changes}.

\changed{As a consequence, the ratio of the irreversibility-driven long-term energy error over the short-term energy error decreases like $\sim\eta^3$ with decreasing $\eta$, independent of the order of the integrator (which determines the short-term energy error). As a consequence, the effects of irreversibilities on the long-term energy error are hardly relevant for sufficiently small $\eta$. Indeed, there is no visible correlation between the instances of irreversible step-size changes in the bottom panel of Fig.~\ref{fig:intg:kepler:dE} (indicated by thin red lines on top and bottom) and the run of the long-term energy error.} This suggests that the methods of Section~\ref{sec:diff} may well be useful in long-term integrations of planetary system dynamics.

%%%%%%%%%%%%%%%%%%%%%%%%%%%%%%%%%%
\section{Conclusion}
\label{sec:conclude}
My attempts to improve on the simple forward method for adapting individual particle step sizes were met with varied success. When merely trying to reduce \changed{deviations form} reversibility (section~\ref{sec:reduce:R:err}), progress is possible either by extrapolating the step size or by a try-and-reject approach, which however comes at increased computational costs. \changed{Porting these methods to $N$-body simulations requires a well-behaved time-step function (without short-term fluctuations), which may require adaptive force softening.}

The situation is more complicated when attempting explicit reversibility. Adapting the scheme of \cite{HolderLeimkuhlerReich2001} to the block-scheme step (section~\ref{sec:explicit}) obtains methods that suffer from oscillations of the step size or of an auxiliary continuous step-size variable. These oscillations are already present in the original method \changed{of \citeauthor{HolderLeimkuhlerReich2001}}, but are much more problematic with the block-step scheme, where they cause irreversible step-size changes. More promising is the idea\changed{, pursued in section~\ref{sec:diff},} to integrate a continuous step-size variable, analogous to the method of \cite{HairerSoderlind2005} \changed{for non-discrete $h$}. This avoids the oscillations and obtains \changed{near-reversibity}, but requires the computation of the time derivative of the time-step function. Such a method may well be useful in long-term integration of planetary systems, when reversibility is much more important than in any other $N$-body model.

It appears that no practical (explicit or nearly explicit) scheme exists to adapt individual block-step discretised particle step sizes \emph{exactly} reversibly. As outlined in sub-section~\ref{sec:reversible?}, the ultimate reason \changed{appears to be} the discretisation of the step size itself.

%%%%%%%%%%%%%%%%%%%%%%%%%%%%%%%%%%%%%%%%%%%%%%%%%%%%%%%%%%%%%%%%%%%%%%%%%%%%%%
\section*{Acknowledgements}
I thank David M.~Hernandez for a critical reading and Scott Tremaine for useful conversations on symplecticity and time symmetry. WD is partly supported by STFC grant ST/N000757/1.
%%%%%%%%%%%%%%%%%%%%%%%%%%%%%%%%%%%%%%%%%%%%%%%%%%%%%%%%%%%%%%%%%%%%%%%%%%%%%%
\bibliographystyle{mnras}
\bibliography{refs}

%%%%%%%%%%%%%%%%%%%%%%%%%%%%%%%%%%%%%%%%%%%%%%%%%%%%%%%%%%%%%%%%%%%%%%%%%%%%%%
\appendix
\section{A flipping scheme}
%%%%%%%
\begin{figure}
	\hfill\includegraphics[width=84mm]{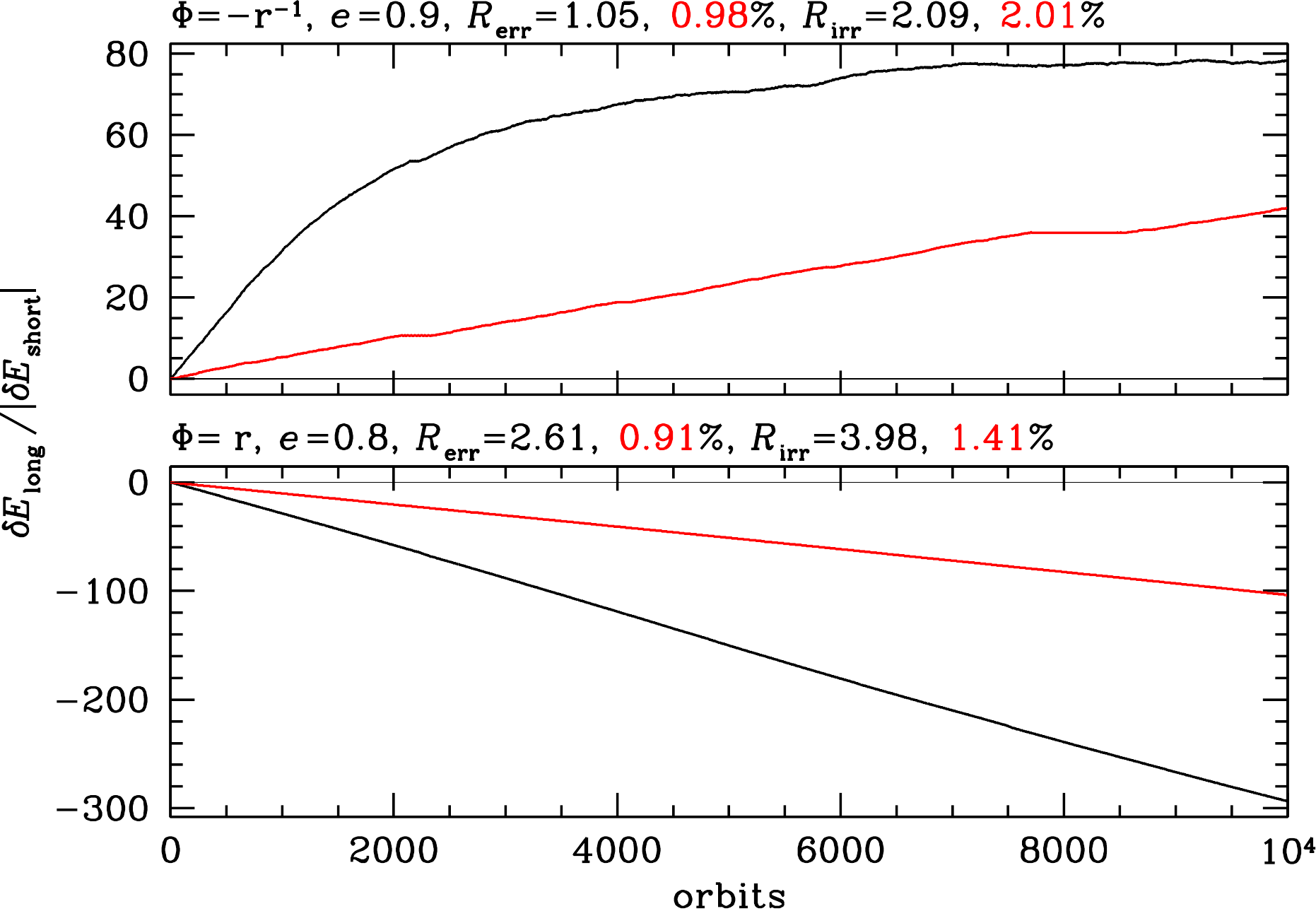}\hfill
	\caption{\label{fig:flip:dE}
	As Fig.~\ref{fig:fwrd:dE} but for the scheme~\eqref{eq:flip}. $R_{\mathrm{irr}}$ is the net rate of irreversible step-size changes per step-size change, defined in equation~\eqref{eq:Rirr}.
	}
\end{figure}
%%%%%%%
The obvious way to adapt the \changed{adaptation method~\eqref{eq:explicit:mu}} to discrete step sizes is (with $\mu$ the geometric mean) to set
\begin{equation} \label{eq:explicit:hnew:failed}
	h_{n+1/2} = h_{\mathrm{block}}(T_n^2/h_{n-1/2}),
\end{equation}
which gives the scheme
\begin{equation} \label{eq:flip}
	\delta r = \left\{\begin{array}{lllcll}
	-1 \;\;& \text{if} \;\; & h_{n-1/2} & \le & \tfrac{1}{\sqrt{2}} T_n & 
	\;\; \text{and block-step allows}, \\[0.5ex]
	\phm0& \text{if} & h_{n-1/2} & \le & T_n,
	\\[0.5ex]
	+1 & \multicolumn{4}{l}{\text{otherwise}}.
	\end{array}\right.
\end{equation}
The only difference to the forward scheme~\eqref{eq:fwrd} is the factor $1/\sqrt{2}$ instead of $1/2$ in the first clause.

%%%%%%%
\begin{figure}
	\begin{center}
		\includegraphics[width=79mm]{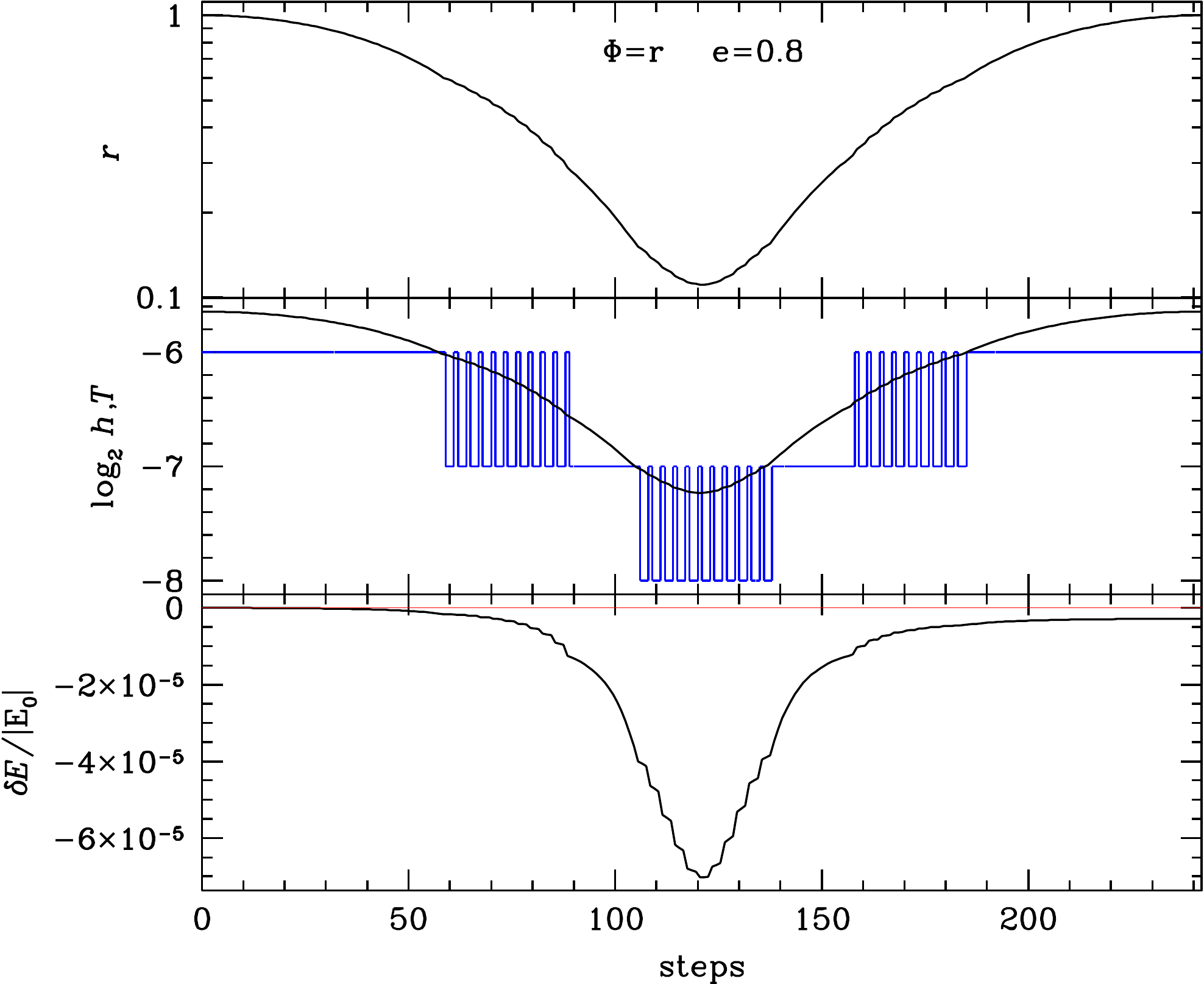}
	\end{center}
	\vspace*{-3mm}
	\caption{\label{fig:flip:orbit}
	As the bottom panel of Fig.~\ref{fig:fwrd:orbit} but for scheme~\eqref{eq:flip} with \changed{$\eta=0.02$ and} long-term energy error reported in the bottom panel of Fig.~\ref{fig:flip:dE}.
	}
\end{figure}
%%%%%%%
Fig.~\ref{fig:flip:dE} shows the \changed{ratio of long- to short-term} energy error for this scheme. Obviously, this is only \changed{a little} better than the forward scheme~\eqref{eq:fwrd}. So obviously constraining the time steps to follow the block-step breaks the reversibility of this method, but how?

Fig.~\ref{fig:flip:orbit} plots the run of radius, $h$, $T$, and energy over the first orbit in the cusp model. One sees first that the step sizes flip between two values if $2^{1/2-r}\le(T/h_{\max})\le2^{1-r}$ for some rung $r$. By itself this flipping is reversible. However, there are ten such flips between steps 59 and 88 on the way to peri-centre, but only nine between steps 158 and 185 on the way out again -- a violation of time symmetry. Upon closer inspection, one finds that the last flip on the in-falling part of the orbit (at step 88) is not reversible: in reversed time the scheme~\eqref{eq:flip} would not chose to change $h=2^{-7}$ to $h=2^{-6}$ at that moment.

\changed{This is a consequence of} the restriction to changes in $h$ by at most a factor two. If allowing larger rung changes, the algorithm jumps from $h=2^{-6}$ to $h=2^{-8}$ at step 88 and in reversed time jumps back. Unfortunately, enabling \changed{$|\delta r|>1$} allows the occasional much too long time step with disastrous consequences for the integration accuracy. \changed{Using} different limits for $\delta r$ and\changed{/or} functions $\mu(x,y)$ \changed{can give} some improvement over the reported attempt, \changed{but did} not yield a reliable method significantly better than the simple forward scheme~\eqref{eq:fwrd}. Moreover, the flipping of step sizes is inefficient for $N$-body force solvers, since in some places half the particles will have rung $r$ and the other $r+1$, switching every other step.

The scheme~\eqref{eq:flip} only produces irreversible step-size adaptations of the first two types in equation~\eqref{eq:Rirr}, which give energy errors of the same sign but opposite to the second two types. 
%%%%%%%%%%%%%%%%%%%%%%%%%%%%%%%%%%%%%%%%%%%%%%%%
\label{lastpage}
\end{document}